\documentclass{elsart}
\usepackage{longtable}
\usepackage{epsfig}
\begin{document}

\begin{frontmatter}

\title{Complementary optical--potential analysis of $\alpha$-particle 
elastic scattering and induced reactions at low energies}

\author{M. Avrigeanu\corauthref{ca}}\ead{mavrig@ifin.nipne.ro},
\author{A.C. Obreja},
\author{F.L. Roman},
\author{V. Avrigeanu}
\address{"Horia Hulubei" National Institute for Physics and 
	Nuclear Engineering, P.O. Box MG-6, 76900 Bucharest, Romania}
and
\author{W. von Oertzen}  
\address{Freie Universit\"at Berlin, Fachbereich Physik, 
	Arnimallee 14, 14195 Berlin, and Hahn--Meitner--Institut, 
	Glienicker Strasse 100, 14109 Berlin, Germany}
\corauth[ca]{Corresponding author. Tel.: +40-21-4042300}

\begin{abstract}
A previously derived semi--microscopic analysis based on the 
Double Folding Model, for $\alpha$-particle elastic scattering on 
$A\sim$100 nuclei at energies below 32 MeV, is extended to medium 
mass $A\sim$50--120 nuclei and energies from $\sim$13 to 50 MeV. 
The energy--dependent phenomenological imaginary part for this 
semi--microscopic optical model potential was obtained including the 
dispersive correction to the microscopic real potential, and used 
within a concurrent phenomenological analysis of the same data basis. 
A regional parameter set for low--energy $\alpha$-particles entirely 
based on elastic--scattering data analysis was also obtained for 
nuclei within the above--mentioned mass and energy ranges. 
Then, an ultimate assessment of $(\alpha,\gamma)$, $(\alpha,n)$ and 
$(\alpha,p)$ reaction cross sections concerned target nuclei from 
$^{45}$Sc to $^{118}$Sn and incident energies below $\sim$12 MeV. 
The former diffuseness of the real part of optical potential as well 
as the surface imaginary--potential depth have been found responsible for 
the actual difficulties in the description of these data, and modified
in order to obtain an optical potential which describe equally well 
both the low energy elastic--scattering and induced--reaction data 
of $\alpha$-particles.
\end{abstract}

\begin{keyword}
$\alpha$--$^{50}$Ti, $^{51}$V, $^{50,52,53}$Cr, $^{56,58}$Fe,
$^{59}$Co, $^{58,60,62,64}$Ni, $^{63}$Cu, $^{89}$Y, $^{90,91}$Zr,
$^{92,94}$Mo, $^{107}$Ag, $^{112,116,122,124}$Sn optical potentials;
Elastic scattering, E$<$50 MeV; 
$^{56}$Fe, $^{58,62,64}$Ni, $^{63}$Cu, $^{70}$Ge, $^{96}$Ru, $^{106}$Cd, 
and $^{112,118}$Sn $(\alpha,\gamma)$, 
$^{45}$Sc, $^{46,48}$Ti, $^{51}$V, $^{50}$Cr, $^{55}$Mn, $^{54}$Fe, 
$^{59}$Co, $^{62,64}$Ni, $^{63,65}$Cu, $^{96,98}$Ru, $^{106}$Cd,
and $^{118}$Sn $(\alpha,n)$, 
$^{44}$Ti, $^{58,62}$Ni, $^{96}$Ru and $^{112}$Sn $(\alpha,p)$, 
E$<$12 MeV; Statistical compound--nucleus reactions
\PACS 21.30.Fe \sep 21.65.+f \sep 24.10.Ht \sep 24.60.Dr \sep 
25.55.Ci \sep 27.60.+j
\end{keyword}
\end{frontmatter}

\tableofcontents

\section{Introduction}

The $\alpha$--nucleus optical model potential (OMP) plays a key role 
in studies of nuclear structure and nuclear reactions, e.g. it is 
used to unify the bound and scattering $\alpha$-particle states 
\cite{peh90}, to analyze the superheavy nuclei $\alpha$--decay 
half--lives \cite{pm06}, in basic nuclear astrophysics applications 
\cite{mar07} and for the estimation of radiation--damage effects in 
fusion test facilities and accelerator--driven systems. A successful 
global optical potential was formerly obtained by Nolte et al. 
\cite{mn87} only for $\alpha$-particle energies above 80 MeV. At 
the same time, it has already been observed that the results 
obtained from the analysis of the low--energy elastic scattering 
data suffer from discrete and continuous ambiguities in the OMP 
parameters. Moreover, two main questions are still open: (i) the OMP 
parameter sets obtained from $\alpha$-particle elastic scattering 
at high energies \cite{mn87} do not describe either the lower--energy 
($<$40 MeV) elastic scattering or the complete fusion data, and 
(ii) the statistical $\alpha$-particle emission is underestimated 
by the OMPs that account for elastic scattering on the ground--state 
nuclei \cite{lwp74,pps76,jma90,va94}. In the latter case, the need 
for new physics in potentials to describe nuclear de--excitation 
within the statistical model calculations has already been pointed 
out \cite{glr87}. Similarly, the effects caused by changes of the 
nuclear density at a finite temperature have been considered in this 
respect within the real part of the double folding model (DFM) of 
the alpha--nucleus optical potential \cite{ma06a}. 

Three improved semi--microscopic global optical potentials have been 
derived recently \cite{pd02} in order to reproduce $\alpha$-particle 
elastic scattering as well as $\alpha$--induced and $(n,\alpha)$ 
reaction data, by using the same DF real potential and WS imaginary 
parts with either a purely volume imaginary term (I), or a volume 
plus surface imaginary potential (II), as well as a damped surface 
potential together with the dispersive contribution to the real DF 
potential (III). In spite of these quite distinct assumptions, it 
has been shown that these three OMPs lead to cross sections which 
do not exhibit any substantial differences apart for some cases at 
backward angles. However, all three of them show an uncertainty 
factor of 10 up to which it has been possible to predict globally 
$\alpha$--induced reaction cross sections \cite{pd02}. As a result, 
in order to avoid any question concerning the remaining parameters 
largely needed within statistical model calculations, a DFM--based 
semi--microscopic analysis of only $\alpha$-particle elastic 
scattering on $A\sim$100 nuclei at energies from $\sim$14 to 32 MeV 
has been carried out \cite{ma03}. The use of this potential at even 
lower energies has provided a suitable description of the $(\alpha,n)$ 
reaction cross sections for lighter target nuclei with $A$$\leq$54 
\cite{ma07a}, while it has led to a major overestimation of 
$(\alpha,\gamma)$ reaction cross sections for $^{106}$Cd \cite{gg06} 
and $^{112}$Sn \cite{no07}. Better results have been provided in the 
later cases by either the well--known mass-- and energy--independent 
four--parameter global potential of McFadden and Satchler \cite{lmf66} 
obtained by analysis of 26 MeV $\alpha$-particle elastic scattering, 
or with the potential of Refs. \cite{tr03,pek04} which was fitted to 
$(n,\alpha)$ and $(\alpha,\gamma)$ reaction data around $A\sim$145. 
At the same time the measured $(\alpha,n)$ and $(\alpha,p)$ reaction 
cross sections \cite{gg06} have been described only by the OMPs of 
Refs. \cite{ma06a,lmf66} while even the $\alpha$--potential of 
Galaviz et al. \cite{dg05}, which was deduced from the 
$\alpha$-particle elastic scattering on $^{112}$Sn at energies close 
to the Coulomb barrier, only poorly  describes the data for 
$\alpha$--capture on $^{112}$Sn at energies lower than 2--6 MeV. 
A common final assumption \cite{gg06,no07}, related to a similar 
overestimation of the $(\alpha,\gamma)$ data and an underestimation 
of the $(\alpha,p)$ data, has been that these deviations are not only 
caused by the $\alpha$--potential. However, the quotation in Ref. 
\cite{gg06} of Ref. \cite{ma06a} for the possible difference between 
optical potential derived from scattering and reaction data, 
discarded the distinction between the incident and the emitted 
$\alpha$-particles discussed within the latter paper. One may 
indeed keep in mind the difference in energy range within which an 
$\alpha$-particle potential is usually established by analysis of 
the elastic scattering, and that of the $\alpha$--induced reactions 
of astrophysical interest. Similar questions have been raised by a 
rather recent analysis for the target nuclei $^{63}$Cu \cite{msb05}, 
$^{96,98}$Ru \cite{wr02} and $^{118}$Sn \cite{sh05}.

A global $\alpha$--nucleus optical potential has been proposed by 
the BARC group \cite{ak06} for $A\sim$12--209 and energies from 
the Coulomb barrier up to about 140 MeV, based on the systematics 
of the real and imaginary potential volume integrals established 
by Atzrott et al. \cite{ua96} by using a DFM real part and the 
description in a unified way of elastic scattering data for 
$A$=40--208 and energies from 30 to 150 MeV, as well as bound
state properties. The BARC global potential has been found to 
describe well the high energy elastic--scattering data, while at 
lower energies the calculations and the data differ considerably 
and further investigation has been found necessary \cite{ak06}. 
Next, an acceptable quality of fits has been considered for the 
calculation of nuclear reactions with $\alpha$-particles 
especially in the entrance channel. However, this conclusion 
could be influenced, at very low incident energies, by the 
comparison with the results obtained by using improperly the 
distinct OMP for emission of $\alpha$-particles \cite{va94}.

The recent high precision measurements of $\alpha$-particle 
elastic--scattering, e.g. \cite{dg05,zf01,ggk07}, pointed to 
additional features of the $\alpha$-particles scattering at low 
energy. Consequently, further eventual improvement of global OMP 
parameters obtained previously through semi--microscopic analysis 
of the low--energy $\alpha$-particle elastic scattering become 
possible \cite{ma06b}. It is discussed in the present work with 
reference to the mass region 50$\leq$$A$$\leq$124 and energies 
below 50 MeV. Following the above--mentioned studies which aim 
to describe both the $\alpha$-particle elastic scattering and 
reaction data, we have first attempted to understand the failure 
of $\alpha$-particle OMPs from elastic--scattering analysis, to 
describe the reaction data. Thus, getting an insight on moving 
below the Coulomb barrier from the energy range where the 
$\alpha$-particle OMPs are usually established by elastic--scattering 
analysis, the eventual difference of $\alpha$-particle potentials 
in the entrance/exit channels \cite{va94,ma06a} could be made clear.

The present work aims firstly to extend the previous semi--microscopic 
analysis of Ref. \cite{ma03} on lighter nuclei ($A\sim$60) in order 
to use the corresponding larger data basis of the $\alpha$-particle 
elastic scattering. The basic model ingredients are given in Sec. 2 
while the results of the semi--microscopic analysis of the experimental 
$\alpha$-particle elastic scattering on $A<$124 nuclei at energies 
below 50 MeV, performed in order to adopt a proper energy--dependent 
phenomenological imaginary part, are described in Sec. 3. The work is 
completed by a full phenomenological analysis of the same data, leading 
to a regional optical potential (ROP) parameter set. Its connection 
to a survey of $\alpha$-particle induced reaction data below 12 MeV is 
given in Sec. 4, starting with the accurate total--reaction cross 
sections of Vonach et al. \cite{hv83} and including the open questions 
recently noted for $^{63}$Cu \cite{msb05} and $A\sim$100 nuclei
\cite{gg06,no07,wr02,sh05}. As a matter of fact, in order to calculate 
the reaction cross section we have used a consistent parameter set 
established by analyzing various independent experimental data for all 
stable isotopes of V, Mn, Co, Ni, Cu \cite{pr01,vs04,ma07b}, Mo 
\cite{pr05}, Pd, Sn and Te \cite{va07}. Thus, it became possible to 
focus on the uncertainties of the $\alpha$-particle OMP parameters 
and their improvement. Final conclusions are provided in Sec. 5. 
Preliminary results have been presented elsewhere \cite{ma07a,ma07c}.
 
\section{The semi--microscopic and phenomenological optical 
potentials}
\subsection{The Double--Folding real potential}

Various attempts have been made to replace the phenomenological 
real potential of Woods--Saxon (WS) type by a microscopic 
$\alpha$--nucleus potential using an effective nucleon--nucleon 
($NN$) interaction, in order to avoid too much phenomenology in 
the description of the $\alpha$-particle elastic scattering  
data. Actually the DF method \cite{grs79}, with an effective 
$NN$--interaction folded with the mass distributions of both the 
target nucleus and the projectile, has been widely used to 
generate OMPs for nucleons, $\alpha$-particles and heavy--ions 
(e.g., \cite{grs79,meb97,dtk01}). The M3Y--Reid \cite{gb77} and 
Paris \cite{na83} are the most familiar interactions 
\cite{grs79,meb97,dtk00}, while a density dependence of the 
$NN$--interaction has been also incorporated \cite{dtk01,dtk93,dtk97}.
However, the M3Y--interactions can be used only to obtain the 
real potential, and the imaginary term must be parameterized 
independently (e.g. \cite{dtk00}) or simply taken from a 
phenomenological OMP \cite{grs79}. As a first result, the 
former approach may reduce the number of the OMP parameters 
and corresponding uncertainties, its success being proved in the 
description of the elastic scattering of many systems \cite{dtk00}. 

More recently the DF formalism for the $\alpha$--nucleus optical 
potential has been revised at $\alpha$-particle energies above 80 
MeV, in order to study the exchange effects and density dependence 
of the effective $NN$--interaction \cite{dtk01}. However, the 
situation is considered less clear for the $\alpha$--nucleus OMP 
at low energies, where the imaginary--potential is strongly energy 
dependent and nuclear structure effects should be taken into 
account, the data being mainly sensitive to the potential at 
the nuclear surface. Therefore, in addition to the results of 
Khoa \cite{dtk01}, suitable constraints for OMP parameters at 
$\alpha$-particle energies around the Coulomb barrier have been 
checked \cite{ma03} using the most advanced DFM version with the 
explicit treatment of the exchange component and no adjustable 
parameter or normalization constant.

The key ingredients of the DFM calculations which are the effective 
$NN$--interaction and the nuclear--density distributions of the 
interacting nuclei, have been enlightened by the analysis \cite{ma03} 
of the $\alpha$--$\alpha$ elastic--scattering angular distributions 
measured at incident energies below the reaction threshold of 34.7 
MeV for the first open channel $^7$Li + p. These data were found to 
be better described by the M3Y--Reid \cite{gb77} than by the 
M3Y--Paris \cite{na83} effective $NN$--interaction, both folded with 
the Baye et al. \cite{db96} $\alpha$-particle density distribution. 
This choice provides a better agreement with the data as compared to 
the Satchler--Love \cite{grs79} and Tanihata et al. \cite{it92} 
forms. Moreover, the density dependence of the M3Y effective 
$NN$--interaction, which accounts for the reduction of the 
interaction strength with increasing density was chosen with a 
linear energy dependence (BDM3Y type) \cite{dtk93} provided by the 
analysis  \cite{ma03} of angular distributions of the elastically 
scattered $\alpha$-particles on $^{90}$Zr at energies between 21 
and 25 MeV.

Finally, the nuclear density distribution of the target nuclei needed 
in the DFM has been described by means of a two--parameter Fermi--type 
function with the parameters chosen to reproduce the electron 
scattering data \cite{jwn70,jwl76} and the shell model calculations 
\cite{mef85}. The basic formulas for the calculations of the DFM 
real part of the optical potential as well as the rest of the model 
assumptions are given in Ref. \cite{ma03}.

\subsection{The semi--microscopic and phenomenological optical 
potentials}

Similarly to the semi--microscopic analysis in Ref. \cite{ma03} of 
the $\alpha$-particle elastic scattering on $A\sim$100 nuclei, at 
energies below 32 MeV, we obtained the energy--dependent 
phenomenological imaginary potential using the corresponding 
dispersive correction $\Delta U(r,E)$ to the microscopic "parameter 
free" DF real potential $U_{DF}(r,E)$ within the complex optical 
potential $U(r)$,

\begin{eqnarray} \label{eq:1}
 U(r) &=& V_C(r) + U_{DF}(r,E) + \Delta U(r,E) + 
 					\: iW_V\: f(r,R_V,a_V) \:
        \nonumber \\
      & & + i W_D\: g(r,R_D,a_D) \: .
\end{eqnarray}
The same volume ($V$) and surface ($D$) imaginary potentials have 
then been involved within a phenomenological analysis of the same 
data basis, leading to a ROP with Woods--Saxon (WS) form factors for 
both real and imaginary parts: 

\begin{eqnarray} \label{eq:2}
 U(r) &=& V_C(r)+V_R\: f(r,R_R,a_R) \: + \: iW_V\: f(r,R_V,a_V) \:
        \nonumber \\
      & & + i W_D\: g(r,R_D,a_D) \: ,
\end{eqnarray}
where
     $f(r,R_i,a_i)$=(1+exp[(r-$R_i$)/$a_i$])$^{-1}$, 
     $g(r,R_i,a_i)$=-4$a_i$d/dr$[f(r,R_i,a_i)]$, and
     $R_i$=$r_i$ $A^{1/3}$, $A$ being the target--nucleus mass number.
     $V_C(r)$ is the Coulomb potential of a uniformly charged sphere 
     of radius $R_C$ while $r_C$= 1.30 fm.
The additional surface term of the imaginary part, with the 
derivative shape, has been introduced at lower energies because 
of the importance of direct reactions, while the number of direct 
channels is still small in this energy range. $W_D$ decreases with 
increasing energy and vanishes at few tens of MeV, e.g. Refs. 
\cite{pd02,zf01,amk82}.

It should be emphasized that no adjustable parameter or normalization 
constant has been involved within this analysis for the real part of
Eq. (\ref{eq:1}) in order to determine the imaginary part of the OMP, 
so that the predictive power of this semi--microscopic potential is 
preserved. On the other hand, the widely used renormalization factor 
\cite{dtk01} for the real semi--microscopic potential is a convenient 
way to take into account the absorption from the elastic channel and 
the presence of couplings to other channels \cite{ha93,ai00}, while 
these effects can be represented by the imaginary--potential dispersive 
contribution to the total real potential in addition to the real DF 
potential \cite{cm86a,cm86b,grs91}. The analytical solution for the 
dispersion relation  \cite{cm86a} adopted and discussed previously 
\cite{ma03} has been used also in the present work.

\section{($\alpha$,$\alpha$) semi--microscopic and phenomenological 
 analysis}

Since the previously published phenomenological analyses of 
$\alpha$-particle elastic scattering data for 50$\le$$A$$\le$124 
target nuclei were performed for various target nuclei only at 
specific incident energies, and the systematic behavior of the mass 
or energy dependences of the corresponding OMP parameters were not
considered, we have looked for a consistent parameter set able to 
describe the bulk of these data at lower, e.g. $\le$50 MeV, incident 
energies. Thus we have analyzed experimental angular distributions of 
$\alpha$-particle elastic--scattering on target nuclei from $^{50}$Ti 
to $^{124}$Sn and $\alpha$-particle energies from 8.1 to 49 MeV (see
Table 1, where an overview on the present data and results of the 
analysis is given). 

Within our two--step OMP approach \cite{ma03}, we determined first the 
parameters of an energy--dependent phenomenological imaginary part 
while the parameter--free DF potential, Eq. (\ref{eq:1}) was used 
for the OMP real part. The computer code SCAT2 \cite{scat2} has been 
used, modified to include the semi--microscopic DF potential of Refs. 
\cite{dtk93} as an option for the real potential. Unfortunately an 
analysis of the $\chi^2$--deviation per degree of freedom between the 
experimental and calculated cross sections, which would have been the 
optimal procedure, has not been possible in all cases due to the lack 
of numerical cross--sections including the corresponding errors for 
some of  the experimental data given in Table 1. Nevertheless, this 
procedure was applied for the more recent data using the original 
errors and a good overall agreement was obtained for various target 
nuclei, thus providing a suitable validation of the actual OMP 
parameter sets. 

In order to have the usual WS--parameterizations according to 
Eq. (\ref{eq:2}), phenomenological analysis of the same data was then 
carried out keeping the imaginary part unchanged from the former 
semi--microscopic analysis. In fact, minor adjustment were involved 
for the imaginary potential depths (Table 2) but none for the related 
average geometry parameters given in Table 3. The advantage of having 
already well settled about half of the usual OMP parameters increases 
obviously the accuracy of the local fit of data (Figs. 1--7). This
procedure reduces to a great extent the problem of OMP continuous 
ambiguities in the real potentials of WS shape, as noted previously 
also by Mohr et al. \cite{pm97}. However, the question of discrete 
ambiguities known as the "family problem", with various real potential 
depths $V_R$ leading to comparable fits to the experimental data as 
shown in Fig. 8 is still open. The similar $\chi^2$ minima have been 
obtained by continuous variation of the $V_R$ value and readjusting 
correspondingly the real--potential geometry parameters. For the 
last two quantities continuous ambiguities within the same discrete 
ambiguity (as, e.g., in Fig. 1 of Ref. \cite{lmf66}) are not shown but 
the values corresponding to the optimum values of $V_R$, in order to 
point out the trend of their change with "families". As a matter of 
fact, the origin of the discrete ambiguity has been well understood 
from the beginning as different numbers of half--wavelengths are 
included within the potential well \cite{lmf66}, while the elastic 
scattering analysis alone is not able to solve this problem 
\cite{rmd63}. However, Mohr et al. \cite{pm97} have shown that, once 
very accurately measured scattering data became available, one may 
discriminate between discrete values of the real--potential volume 
integrals per interacting nucleon pair, given by the general form

\renewcommand{\theequation}{\arabic{equation}}
\begin{equation} \label{eq:3}
 J_x(E)=\frac{1}{A_1\:A_2}\int U_x(r,E) \: d^3r .
\end{equation}

\noindent
where $x$=$R,V,D$. Thus we have looked also for the $J_R$ values 
corresponding to the $\chi^2$ minima in Fig. 8 as well as for all 
angular distributions analyzed in the present work (Table 2). 
As a result, most of these minima are related to values of about 
$J_R\sim$380--440 MeV$\:$fm$^3$, around the $\alpha$-particle energy 
of 25 MeV (Fig. 1), slightly decreasing with energy. Therefore 
these best--fit WS real--potential parameters have been selected 
in the cases where more similar $\chi^2$ minima exist, and 
involved subsequently in a procedure of deriving average mass--, 
charge--, and energy--dependent parameters for the mass and energy
ranges of this work, but as close as possible to those introduced 
by Nolte et al. \cite{mn87} above 80 MeV. Since it is well--known 
\cite{lmf66}, that a linear interpolation between the optimum values 
does not lead to parameter values which describe the data 
reasonably well, the dotted curves in Fig. 8 having only the role 
to display a trend, these average dependences were not obtained at 
once for all real potential parameters. In order to reduce as much 
as possible the adverse effects of the averaging, the fit of the 
data was repeated each time after the average dependences were 
derived, and the result consequently used first for the reduced 
radius and afterward for the potential depth. The local $a_R$ values 
obtained finally by using average values for the rest of all OMP 
parameters are shown in Fig. 9 versus the $\alpha$-particle energy 
as well as for ratio to the Coulomb barrier value. The mass 
dependence of the values of the interaction radius $R_B$ \cite{wn80} 
used in this respect, given in Table 3, emphasizes a similar 
importance of energy-- and mass--dependences of the real--potential 
diffuseness, appropriately described by the form given in Table 3 
for energies from the Coulomb barrier to $\sim$25 MeV. However, the 
spread of these local parameter values at higher energies has not 
been fully understood, so that an average value of $a_R$=0.55 fm has 
been adopted for the energy range above a value $E_3$ given in Table 3.

While the real $J_R$, volume $J_V$ and surface $J_D$ imaginary
components of the local phenomenological potentials are also given in 
Table 2, the $J_R$ values corresponding to the average OMP parameters 
of Table 3 decrease from $\sim$410 to $\sim$370 MeV$\:$fm$^3$ between 
the $\alpha$-particle energies $\sim$12--50 MeV. These values 
correspond to volume integrals between $\sim$310--280 MeV$\:$fm$^3$ 
for the microscopic DFM potentials which provide a similar description 
to the elastic scattering data at low energies, and being almost 
identical in the tail region (e.g., Fig. 10 of Ref. \cite{ma03}). 
Actually, the differences between these volume integrals of 
microscopic DF and phenomenological real potentials are simply due to
their different radial dependence, which should be taken 
into account when the two types of potentials are compared. Therefore 
we have once more found \cite{ma03} that our $J_R$ values correspond 
to microscopic DFM volume integrals which are 10$\%$ lower than 
expected from the systematics found in \cite{dg05,ua96,zf01,pm97,pm00}. 
They are also similar to those obtained by using the semi--microscopic 
potential III of Ref. \cite{pd02} at the same energies, or derived 
from the analysis of $\alpha$--decay \cite{pm06,pm00}.

The angular distributions calculated using these average parameters 
(regional optical potential -- ROP) are shown in Figs. 1--7, in order to 
emphasize its usefulness and related deviations from the local analysis 
results. Moreover, the predictions of the most recent global parameter 
set of Kumar et al. \cite{ak06} are included, while those provided by 
McFadden and Satchler with a four--parameter global potential 
\cite{lmf66} have already been presented elsewhere \cite{ma07a}. 
A comparison of semi--microscopically calculated angular distributions 
and ROP results is shown in Fig. 10(a) for the lowest $\alpha$-particle energies on $^{58}$Ni, within a discussion to be given in Section 
4.2 on the effects of a ROP particular amendment below the Coulomb barrier.
On the whole, a really improved description of the data is provided by 
the present ROP. Moreover, it should be noted that a rather suitable 
account of the data is also obtained using OMPs with real--potential 
diffuseness $a_R$ \cite{ma03,lmf66} notably lower than the $a_R$ values 
which are needed in order to describe $\alpha$-particle emission from 
excited compound nuclei \cite{va94,ma06a}. 
   
\section{$(\alpha,x)$ and total $\alpha$--reaction cross section 
analysis}

\subsection{Particular ROP features below the Coulomb barrier}

The enlarged analysis of the $\alpha$-particle optical potential 
based on the elastic scattering data, achieved mainly at energies
above the Coulomb barrier, makes one confident to check its 
suitability at even lower energies. Actually one may expect that a 
simple extension of the corresponding OMP parameters, below the 
energies involved within their establishment, could be reliable 
provided that these parameters vary regularly over the entire energy 
range. For the $\alpha$-particle energies below the Coulomb barriers
this could be the case of, e.g., the real and volume imaginary 
potential but not for the surface imaginary potential according to 
the related comment in Sec. 2.2. The analysis of the $(\alpha,x)$ 
reaction cross sections below the Coulomb barrier may reveal OMP 
parameters different by those at the former energies. On the other
hand, a change of the surface imaginary potential depth below the 
Coulomb barrier is related indeed to a variation of the corresponding 
dispersive correction (inset in Fig. 10) to the real part of the 
semi--microscopic potential, so that an additional analysis should
concern the eventual effects on the first step of our ROP setting up. 

First, we have used the ROP parameters given in Table 3 for energies 
above the Coulomb barrier, for Hauser--Feshbach statistical model 
calculations of $(\alpha,x)$ reaction cross sections. They were 
carried out similarly to the previous analysis of $(n,\alpha)$ 
reaction cross section for $A\sim$90 \cite{ma06a,ma95}, except the 
investigated OMP was now related to $\alpha$-particles in the 
incident channel. For the rest of statistical--model parameters we 
have used consistent sets established by analyzing various 
independent experimental data for all stable isotopes of V, Mn, Co, 
Ni, Cu \cite{pr01,vs04,ma07b}, Mo \cite{pr05}, and Pd, Sn and Te 
\cite{va07}. While a suitable description of the $(\alpha,n)$ 
reaction cross sections was found for lighter target nuclei with 
$A$$\leq$54 (Fig. 11), well improved over the $\sim$10\% accuracy 
\cite{ma07a} provided by the global parameter sets of McFadden and 
Satchler \cite{lmf66}, a major overestimation of $(\alpha,\gamma)$ 
reaction cross sections resulted for the target nuclei $^{62,64}$Ni 
(Fig. 12), $^{63}$Cu (Fig. 13), $^{96,98}$Ru (Fig. 14), and 
$^{106}$Cd and $^{112,118}$Sn (Fig. 15). For this latter class of 
nuclei, definitely marked by larger charge and atomic numbers, even 
the global OMP of McFadden and Satchler provided better results 
\cite{gg06,no07,msb05,wr02} than our former ROP based on 
elastic--scattering data analysis \cite{ma03}. On the other hand, 
the calculated reaction cross sections corresponding to this global 
potential vary with respect to the experimental data, from an 
underestimation around 10\% for target nuclei with $A\sim$60 
\cite{ma07a} to much larger overestimation for $A>$100 
\cite{gg06,no07}. Therefore, in order to understand better the 
behavior of these reaction data, we chose to analyze them against 
the ratio $E_{c.m.}/B_C$ between the incident energy in the 
center--of--mass system and the Coulomb barrier (Figs. 11--15). 
The total $\alpha$--reaction cross sections provided only by the ROP 
parameters established by the elastic--scattering analysis alone are 
additionally shown in Figs. 12--15 for the target nuclei with $A>$60. 
The energy ranges within which the $(\alpha,\gamma)$ and subsequently 
$(\alpha,n)$ reactions control the whole $\alpha$-particle 
interaction by the $\alpha$-particle OMP become thus obvious.

A major source of uncertainty has been the real--potential 
diffuseness $a_R$, which is the OMP parameter marked by the largest 
sensitivity of statistical--model calculated cross sections (e.g., 
Ref. \cite{ma06a}). The systematics provided by the analysis of the 
elastic--scattering data for $E_{c.m.}/B_C>$0.9 (Fig. 9) imply a 
real--potential diffuseness increasing with the energy decrease, 
confirmation of this behavior being essential since a similar trend 
is specific to OMP describing the $\alpha$-particle emission 
\cite{va94,ma06a}. Its establishment at very low energies would 
therefore override any additional assumption related to different 
OMP parameter sets in the incident and emergent channels, respectively. 
However, the account of measured $(\alpha,x)$ reaction cross sections 
seems to presume a decrease of this parameter with the energy decrease. 
Thus, with no other data at hand, a constant $a_R$ value appears as
the foremost option for the real--potential diffuseness at 
energies lower than a value, $E_2$, corresponding to 0.9$B_C$ (Table 3). 
A second run of statistical--model calculations using this assumption 
shown by dashed curves for some of the reactions in Figs. 12--15 
provides calculated cross sections lower than the former ones by less 
than 10\% for $A\sim$60 (Fig. 12) and $\sim$15\% for $A>$100 (Fig. 15). 
This minor change proves that a constant $a_R$ below the Coulomb barrier, 
being yet an appropriate choice within the actual knowledge, can not 
account for the low energy $(\alpha,x)$ reaction data.

Since the real and volume imaginary potential are less uncertain, as 
noted in the beginning of this section, the surface imaginary potential 
remains the central point of discussion. Actually Mohr et al. \cite{pm97} 
pointed out that the strong change of the number of open reaction 
channels close to the Coulomb barrier leads to a strong variation of the 
surface and volume imaginary potentials. This variation has been described 
by a parametrization of the imaginary--potential volume integral 
either according \cite{ua96,pm97} to Brown and Rho \cite{geb81} or by
using a Fermi--type function \cite{pd02,zf01,es98}. On the other hand 
it has been pointed out \cite{dg05,zf01} that the actual ambiguities do 
not allow to determine the shape of the imaginary potential and reduce 
the reliability of extrapolations to lower energies. Therefore, our 
final change of the ROP obtained by elastic--scattering analysis, in 
order to describe the $(\alpha,x)$ reaction data at these energies, 
has been the drop of the surface imaginary potential depth with the 
decrease of $\alpha$-particle energy below the energy limit $E_2$. 
This choice is consistent with the above--mentioned strong change of 
the number of open reaction channels close to the Coulomb barrier, 
tightly matched by the $E_2$ values with the mass dependence shown 
in Fig. 16. Following this consideration and the demand of suitable 
description of the excitation functions in Figs. 12--15, we have 
obtained the corresponding slope of the $W_D(E)$ energy--dependence 
in Table 3. Its average value of 6 would then determine the starting 
point $E_1$ of the surface imaginary--potential energy range. However, 
in order to avoid overall numerical problems just above reaction 
thresholds, we have not finally chosen a vanishing value of the 
potential depth $W_D$ at the very low energy but a minimum value of 
4 MeV. This depth limit and aforementioned slope of the $W_D(E)$ 
eventually lead to the $E_1$ global expression given in Table 3 for 
the final form of ROP in the whole energy range below 50 MeV. Other 
$E_1$ values, related to a constant value of $W_D$ lower than 4 MeV 
at the null $\alpha$-particle energy, may even provide an improved 
agreement with the measured data in particular cases mentioned in 
the following. Moreover, one can now explain the poorer results for 
$(\alpha,x)$ reaction data of the former $\alpha$-particle ROP based 
on the elastic--scattering analysis alone \cite{ma03} as compared to 
the four--parameter global potential \cite{lmf66} and related OMPs 
\cite{tr03,pek04} which have only a volume imaginary potential with 
a constant depth. As a result, the lack of an energy--decreasing 
surface component has prevented larger deficiencies in the fits 
which arise in extrapolations below the Coulomb barrier.

\subsection{Effects within the whole energy range of ROP particular 
amendment below the Coulomb barrier}

The latest change of the surface imaginary potential depth below 
the Coulomb barrier should be followed by a corresponding 
change, through the dispersive relations with an integral over all 
incident energies, in the real part of the semi--microscopic potential 
which has been the first step of our ROP setting up. Therefore, one may
consider that the iterative procedure applied in order to find the best
description of the elastic--scattering data should be resumed by using
the modified surface imaginary potential depth within the dispersive 
correction formula, e.g., Eq. (4.2) of Ref. \cite{ma03}. The corresponding
correction to the DF real potential for the target nucleus $^{58}$Ni, due 
to only the final form of the ROP surface imaginary--potential (Table 3), 
is shown within the inset in Fig. 10(a) while the rest of the matching 
discussion is as for Fig. 4 of Ref. \cite{ma03}. This correction is 
compared with the former curve, obtained by using the analytical solution 
and forms adopted previously \cite{ma03,ma06b}, as mentioned in Sec. 3.2. 
The only change in this respect concerned, following the trend of the 
local parameter values found in this work as well as the lack of data below 
$\alpha$-particle energy of $\sim$9 MeV, a constant increasing surface 
imaginary potential depth between the axes origin and the corresponding 
prediction in Table 3 at the latter energy. It results that the difference 
between the former and final values of this particular dispersive 
correction is well decreasing below 2 MeV for energies increasing above 
$\sim$15 MeV. This quantity is however quite low with respect to the real 
potential depths around 50 MeV within the nuclear surface region, where
this correction is effective. At the same time, if this change would 
be involved in a new iteration of the first step of our ROP setting up, 
for the the imaginary potential parameters, it is obvious that the 
consequent changes of these potential depths would be comparable to the 
minor adjustment of the same depths mentioned in Sec. 3, carried out 
within the second step of our approach. Larger changes of $\leq$5 MeV 
for the dispersive correction due to the surface--imaginary potential 
variation below the Coulomb barrier are present only in the case of the 
elastic--scattering data on $^{58}$Ni between 8 and $\sim$10 MeV. 
Nevertheless, it would be less effective also at these energies to look 
for imaginary--potential parameters within a new iteration of the 
semi--microscopic analysis of elastic--scattering data, due to the already 
very low sensitivity of the calculated elastic--scattering cross sections
to the imaginary potential. Thus, except some obvious limits, the 
corresponding $\chi^2$ minima are so slowly varying with the imaginary
potential depths that no real change of these parameters may result. 
Alternatively, we have chosen to compare in Fig. 10(a) the semi--microscopical  
elastic--scattering cross sections on $^{58}$Ni at these low energies 
calculated by using the same local imaginary--potential parameters
(Table 2) but the different former and final dispersive corrections shown 
in the inset as well. Even these effects are small and obviously within 
the experimental data bars. 

Another point which had to be made clear was the consequence of the change itself of ROP parameters $a_R$ and $W_D$ due to the $(\alpha,x)$ reaction 
cross section analysis, on the formerly calculated angular distributions 
of $\alpha$-particle elastic--scattering by using the parameter 
forms established for energies higher than $E_2$ (Table 3). This again has
been the case of only the elastic--scattering data on $^{58}$Ni between 
8 and $\sim$10 MeV, shown in Fig. 10(b) to have very similar description
before and after the above--mentioned changes of the $a_R$ and $W_D$ 
parameter forms. It should be noted once more that this status follows 
the low variation of these parameters at energies rather close to the 
$E_2$ limit, as well as the minor sensitivity to the imaginary potential 
of the corresponding elastic--scattering cross sections dominated by the 
Rutherford component. Hence we may conclude that the changes of the ROP 
parameters below the Coulomb barrier have effects already within 
uncertainties of the parameter values in the rest of the energy range, 
so that the ROP parameters in Table 3 describe equally well both the 
elastic--scattering and induced--reaction data of $\alpha$-particles.

\subsection{$A$$\leq$54 target nuclei} 

The particularly accurate $(\alpha,n)$ and total $\alpha$--reaction 
cross sections for $^{48}$Ti and $^{51}$V measured and respectively 
established by Vonach et al. \cite{hv83}, obviously overestimated by 
the $\alpha$-particle emission OMP \cite{va94}, are slightly better 
described by the final ROP than by its former version obtained through 
the analysis of the elastic--scattering data alone (Fig. 11). The same 
has been proved \cite{ma07a} to be true also for an enlarged 
systematics \cite{ajm92,cmb04,aev74,vyh93,xp99,ajm94,sgt91} discussed 
recently \cite{ak06}, and several data sets \cite{vyh89,aas00} 
lastly taken into account. The related incident--energy range can 
hardly explain the difference with respect to the same comparison 
just above A=60 (Fig. 12). The reference to the Coulomb barrier 
values, so close to the edge $E_2$ (Fig. 16) between the two distinct 
energy regions of the present ROP (Table 3), is ultimately making the 
point clear. The energies concerning these target nuclei are mainly 
above and around the Coulomb barrier, while in the rest of cases they 
are clearly below it. The two calculated functions for each reaction 
are indeed plainly different only for $E_{c.m.}/B_C<$0.9, while their 
variance is less visible along the steep part of the excitation 
functions.

\subsection{$A\sim$63 target nuclei} 

The data for the target nuclei $^{58,62,64}$Ni \cite{aev74,fkmcg64,mr74,jlz79} 
have been important for the aims of the present work since they illustrate 
the cross--sections trend for nearby isotopes of the same element and the
$(\alpha,\gamma)$ as well as $(\alpha,n)$ and $(\alpha,p)$ reactions, 
which represent the total $\alpha$--reaction cross section in the former 
and latter halves of the energy range concerned, respectively (Fig. 12). 
Actually only the data for the $^{62,64}$Ni target nuclei \cite{jlz79}
were formerly involved in the analysis leading to the final ROP parameters
(Table 3), which were consequently applied within the cross--section 
calculation for $^{55}$Mn \cite{sgt93} and $^{58}$Ni. On the other hand,
the best description has been obtained for the smoother data in the 
latter case, due to the related energy range and Q--effects, while 
less accurate results were obtained for the heavier isotopes in Fig. 12. 
Fortunately, this was not the case of the more recent data \cite{msb05}
for the $^{63}$Cu$(\alpha,\gamma)^{67}$Ga reaction. 
The odd proton--numbers of the target and the compound and residual 
nuclei, with larger nuclear level densities, make the statistical--model 
calculations more reliable. The appropriateness of the last comment is 
also proved by the agreement of the calculated and experimental cross 
sections, although the same consistent parameter set \cite{vs04,ma07b} 
has been involved for all four target nuclei. 

A similar case but due to a distinct reason occurs for cross 
sections of the $(\alpha,\gamma)$ reaction on $^{56}$Fe and $^{70}$Ge
obtained as well as for $(\alpha,n)$ reaction (Fig. 13) by using 
straightforwardly the final ROP. While the energy range of the existing 
$(\alpha,n)$ reaction data was as large as for nuclei with $A$$\leq$54
with respect to the Coulomb barrier, that of the $(\alpha,\gamma)$ 
reaction was again quite narrow. However, the corresponding cross sections
have been very close to the total reaction cross sections over at least
the former half of this energy range. It results thus a larger sensitivity 
of the model calculations to the $\alpha$--particle OMP parameters, in
comparison with that of the $(\alpha,n)$ reaction analysis for similar 
nuclei also shown in Fig. 13. Nevertheless, a very good agreement with 
the measured data has been achieved for both reactions. Altogether, the 
analysis for Ni and Cu isotopes has validated the ROP predictions for 
the total $\alpha$--reaction cross section over more than four orders of 
magnitude. 

\subsection{$A$$\geq$100 target nuclei} 

The simultaneous measurements of the $(\alpha,\gamma)$, $(\alpha,p)$, 
and $(\alpha,n)$ reaction cross sections on the target nuclei 
$^{96}$Ru \cite{wr02} and $^{106}$Cd \cite{gg06} provide the most 
useful support for the $\alpha$-particle OMP discussion at the 
lowest energies below the Coulomb barrier. The measured 
data of the first two of these reactions on the target nucleus 
$^{112}$Sn \cite{no07} have rather similar conditions, while the 
related neutron emission threshold is much higher in energy. The 
simultaneous description of these data is most important for the 
validation of $\alpha$-particle OMP below the Coulomb barrier, 
especially within the energy ranges where the $(\alpha,\gamma)$ 
and subsequently $(\alpha,n)$ reactions stand for nearly the whole 
range of the $\alpha$-particle interaction. On the other hand, the 
assumption \cite{gg06,no07} concerning the similar overestimation 
of the $(\alpha,\gamma)$ cross sections and underestimation of the 
$(\alpha,p)$ cross sections, caused not only by the 
$\alpha$--potential, led us to an additional investigation in this 
respect \cite{va07}. Thus, we  analyzed the neutron total cross 
sections for the Cd, Sn and Te isotopes and the neutron energies up 
to 30 MeV, as well as the neutron capture on the same target nuclei 
for the neutron energies up to 3 MeV. A better knowledge of the 
neutron OMP and then of the $\gamma$--ray strength functions has 
been therefore obtained and applied within the present study of the 
$(\alpha,x)$ reaction cross sections.

First, we can state, that the calculated cross sections for the 
$(\alpha,\gamma)$ reaction on $^{96}$Ru (Fig. 14), $^{106}$Cd and 
$^{112}$Sn (Fig. 15) are in good agreement with the measured data, 
especially at $\alpha$-particle energies where this reaction stands 
for nearly the whole range of the $\alpha$-particle interaction. An 
even better description of these data could be obtained with small 
adjustments of the ROP parameters $a_R$ and $W_D$, since our main 
interest is to obtain a general account of more experimental data 
using the same parameter set. The results are less satisfactory for 
the target nucleus $^{118}$Sn (Fig. 15). However, this happens at 
$\alpha$-particle energies where the $(\alpha,\gamma)$ reaction 
cross sections are at least one order of magnitude smaller than 
the total $\alpha$--reaction cross sections which go mainly in the 
$(\alpha,n)$ reaction channel. 

Secondly, the cross sections of the concurrent reactions 
$(\alpha,p)$, and $(\alpha,n)$ on $^{106}$Cd are quite well 
reproduced, as well as the $(\alpha,p)$ reaction on $^{112}$Sn 
(Fig. 15). A particular case is that of the target nucleus $^{96}$Ru 
at the $\alpha$-particle energies around the threshold where the 
neutron emission becomes possible and increases very rapidly, at the 
expense of the $(\alpha,p)$ reaction cross sections. Since this effect 
is observed similarly to the $(\alpha,n)$ reaction on $^{98}$Ru 
(Fig. 14), one may consider this model--calculation failure to be the 
result of a less suitable account of the neutron and proton emission 
channels. This observation also explains the approximately correct 
reproduction of the slope of the $^{118}$Sn$(\alpha,\gamma)^{121}$Te 
excitation function. Nevertheless, the comparison of the calculated 
and measured $(\alpha,\gamma)$ reaction cross sections at energies 
where they represent the total $\alpha$--reaction cross sections
gives a good validation of the $\alpha$-particle ROP.

\section{Conclusions}

The analysis of the $\alpha$-particle elastic scattering on 
50$\le$$A$$\le$124 nuclei at energies below 50 MeV has been carried 
out using the semi--microscopic DF approach. The energy--dependent 
phenomenological OMP imaginary part, which was obtained in this case, 
using the dispersive correction to the DF real potential, has been 
then used for a completely phenomenological analysis of the same data 
basis. By using the real DF potential and adjusting only the imaginary 
OMP part parameters, the number of free parameters are well decreased. 
Afterward, we adjusted only the parameters of the real phenomenological 
potential of Woods--Saxon shape while the imaginary components remained 
unchanged. An average of the local parameter values has been obtained 
in order to provide finally a regional parameter set which can be 
easily used in further analyses or predictions.

Moreover, the ROP obtained in the present work has been applied 
in the statistical--model analysis of the $(\alpha,x)$ reaction cross 
sections below the Coulomb barrier available below 10--12 MeV for 
nuclei within the same mass range. For the rest of statistical--model 
parameters we have used consistent sets established by analyzing 
various independent experimental data for all stable isotopes of 
V, Mn, Co, Ni, Cu \cite{pr01,vs04,ma07b}, Mo \cite{pr05}, and Pd, Sn 
and Te \cite{va07}, which finally allow us to focus on the 
uncertainties of the $\alpha$-particle OMP. 
While a suitable description of the $(\alpha,n)$ reaction cross 
sections was found for lighter target nuclei with $A$$\leq$54, a major 
overestimation of $(\alpha,\gamma)$ reaction cross sections resulted 
for $^{62,64}$Ni, $^{63}$Cu, $^{96,98}$Ru, $^{106}$Cd and 
$^{112,118}$Sn. This behavior is related to the energies above and 
around the Coulomb barrier, for the former mass range, and clearly 
below it for the latter. A suitable description of the $(\alpha,x)$ 
reaction data below the Coulomb barrier is no longer possible by 
means of the optical potential provided by the elastic--scattering 
data analysis above this barrier, a modified surface imaginary 
potential being necessary.
A reliable assessment of the final ROP below the Coulomb barrier has
been given by its straightforward and successful use in calculations 
of reaction cross sections for the target nuclei $^{44}$Ti, $^{55}$Mn, 
$^{56}$Fe, $^{59}$Co, $^{58}$Ni, $^{65}$Cu, and $^{70}$Ge.
 
Actually, the drop of the surface imaginary potential depth with the 
decrease of $\alpha$-particle energy below the Coulomb barrier, 
necessary for the $(\alpha,x)$ reaction data account, is in line with 
the strong change of the number of open reaction channels within this 
energy range. It can also explain the even worse results of the former
$\alpha$-particle ROP based on the elastic--scattering analysis alone
\cite{ma03}, when used in the $(\alpha,x)$ reaction data analysis, as 
opposed to the four--parameter global potential \cite{lmf66} and 
related OMPs \cite{tr03,pek04} which have only a volume imaginary 
potential with a constant depth. The lack of an energy--dependent 
surface component thus prevents larger negative effects by extrapolation 
below the Coulomb barrier. On the other hand, becoming aware of the
changes of $\alpha$-particle OMPs obtained through elastic--scattering 
analysis alone, in order to describe the $(\alpha,x)$ reaction--data as
well, the difference between the $\alpha$-particle OMPs in the entrance 
and exit channels could be easier understood. Thus, the present insight 
into the behavior of the $\alpha$-particle optical potential at very low 
incident energies will make possible a further concluding assessment of 
the $\alpha$-particle emission at similar energies but in the exit 
channels \cite{va94,ma06a}. 

\section*{Acknowledgments}
This work was supported in part by the EURATOM--MEdC Fusion 
Association, the Research Contract No. 12422 of the International 
Atomic Energy Agency, and MEdC (Bucharest) Contract No. CEEX--05--D10--48.

\newpage

\begin{table*}
\caption{\label{Table1}The experimental data of $\alpha$-particle 
elastic scattering analysed in this work.}
\begin{tabular}{llc}\\ \hline \hline
\hspace{17mm}Target &\hspace{23mm} E$_{\alpha}$        & Ref. \\
                    &\hspace{20mm} (MeV)               \\    \hline
$^{50}$Ti & 24.97, 27.06, 29.06, 31, 33, 35.06, 37.12, & \cite{hpg81}\\
          & 39, 41.05, 43, 44.98, 46.8 \\
$^{51}$V, $^{50,52,53}$Cr, $^{59}$Co, $^{63}$Cu & 25 & \cite{fb90} \\
$^{50}$Cr, $^{62}$Ni & 12.8, 14.56, 16.34, 18.13 &     \cite{ab94}  \\
$^{52}$Cr & 23, 24.97, 27.06, 29.1, 31, 33, 35.06, 37.12,&\cite{hpg81} \\
          & 39, 41.05, 43, 44.98, 46.8 \\
$^{56,58}$Fe &                                    25 & \cite{fb89} \\
$^{56}$Fe &                                    26.45 & \cite{bw62} \\
$^{58,60,62,64}$Ni &                              25 & \cite{fb87} \\
$^{58}$Ni &                            8.1, 9.1, 9.6 & \cite{lrg03} \\
$^{58,60,62,64}$Ni &                18, 21, 24.1, 27 & \cite{wt74} \\
$^{58}$Ni &                         26.5, 29, 34, 38 & \cite{ab78} \\
$^{60}$Ni &                                   29, 34 & \cite{ab78} \\
$^{58}$Ni &                               37, 43, 49 & \cite{uk78} \\
$^{58,60,62,64}$Ni &                            32.3 & \cite{aac74} \\
$^{70,72,74,76}$Ge &		 		  25 & \cite{fb88g} \\
$^{76,78,80}$Se &				  25 & \cite{fb88s} \\	
$^{89}$Y, $^{90,91}$Zr &                21, 23.4, 25 & \cite{wit75} \\
$^{90}$Zr, $^{107}$Ag &                           15 & \cite{watson71} \\
$^{92}$Mo                       & 13.83, 16.42, 19.5 & \cite{zf01} \\
$^{94}$Mo, $^{107}$Ag, $^{116,122,124}$Sn &     25.2 &\cite{bespalova92}\\
$^{89}$Y &                               16.2, 19.44 & \cite{ggk07} \\
$^{112}$Sn &                                    14.4 & \cite{dg05}\\
$^{112,124}$Sn &                                19.5 & \cite{dg05}\\
\hline \hline
\end{tabular}
\end{table*}

\clearpage
\begin{longtable}{lccccccccc}
\caption{\label{Table2}
Optical potential parameters and volume integrals (without the 
negative sign) obtained by fits of the $\alpha$-particle 
elastic--scattering data for A=50--124 nuclei at energies $\leq$50 MeV. 
$E_{\alpha}$, $V_R$, $W_V$ and $W_D$ are in MeV, $r_R$ and $a_R$ 
are in fm, $J_R$, $J_V$ and $J_D$ are in MeV$\cdot$fm$^3$.}\\

\hline \hline
Target&$E_{\alpha}$& $V_R$ & $r_R$ & $a_R$ & $W_V$ & $W_D$ & $J_R$ & $J_V$ & $J_D$\\ 
      & (MeV)      & (MeV) &  (fm) &  (fm) & (MeV) & (MeV) &(MeV$\cdot$fm$^3$)&(MeV$\cdot$fm$^3$)&(MeV$\cdot$fm$^3$)\\
\hline
\endfirsthead
\caption{continued.}\\
\hline
Target&$E_{\alpha}$& $V_R$ & $r_R$ & $a_R$ & $W_V$ & $W_D$ & $J_R$ & $J_V$ & $J_D$\\ 
     & (MeV)      & (MeV) &  (fm) &  (fm) & (MeV) & (MeV) &(MeV$\cdot$fm$^3$)&(MeV$\cdot$fm$^3$)&(MeV$\cdot$fm$^3$)\\
\hline
\endhead
\hline
\endfoot
$^{50}$Ti & 24.97 & 111   & 1.498 &0.540& 19.8 & 2.93& 427 & 55.0 & 10.8 \\ 
          & 27.06 &  53.0 & 1.612 &0.522& 22.2 &  0  & 250 & 61.4 &  0   \\ 
          & 29.06 & 138   & 1.452 &0.552& 24.4 &  0  & 489 & 67.6 &  0   \\ 
	  & 31.0  & 101   & 1.482 &0.572& 26.5 &  0  & 381 & 73.6 &  0   \\ 
	  & 33.0  & 106.5 & 1.468 &0.572& 28.8 &  0  & 392 & 79.7 &  0   \\ 
	  & 35.06 & 109.5 & 1.474 &0.558& 31.0 &  0  & 405 & 86.1 &  0   \\ 
	  & 37.12 &  89.5 & 1.514 &0.550& 33.3 &  0  & 356 & 92.4 &  0   \\ 
	  & 39.0  &  91.5 & 1.510 &0.546& 35.4 &  0  & 361 & 98.2 &  0   \\ 
	  & 41.05 &  92.0 & 1.518 &0.540& 37.7 &  0  & 368 & 105.0&  0   \\ 
	  & 43.0  &  99.5 & 1.502 &0.532& 39.9 &  0  & 385 & 111 &  0    \\ 
	  & 44.98 & 106   & 1.500 &0.524& 42.0 &  0   & 408 & 117 &  0    \\ 
	  & 46.8  & 112   & 1.492 &0.516& 44.1 &  0   & 423 & 122 &  0    \\ 
$^{51}$V  & 25.0  & 115   & 1.476 &0.562& 19.8 & 3.0  & 427 & 54.8 & 10.9 \\ 
$^{50}$Cr & 12.8  & 188   & 1.236 &0.700&  6.3 &20.5 & 458 & 17.6 & 75.4 \\ 
	  & 14.56 & 105   & 1.536 &0.568&  8.3 &18.0 & 438 & 23.0  & 66.0 \\ 
	  & 16.34 & 136   & 1.328 &0.660& 10.3 &15.4 & 393 & 28.4 & 56.6 \\ 
	  & 18.13 & 103   & 1.466 &0.612& 12.2 &12.8 & 383 & 34.0 & 47.1 \\ 
	  & 25.0  & 109  & 1.506 &0.562& 19.9 & 2.9 & 429 & 55.1 & 10.6 \\ 
$^{52}$Cr & 23.0  & 103.5 & 1.516 &0.538& 17.5 & 6.0 & 411 & 48.4 & 21.6 \\ 
	  & 24.97 & 106   & 1.514 &0.524& 19.7 & 3.1 & 418 & 54.6 & 11.2 \\ 
	  & 27.06 & 113.5 & 1.510 &0.512& 22.0 & 0.1 & 442 & 60.9 & 0.45 \\ 
	  & 29.1  & 119.5 & 1.504 &0.506& 24.3 &  0   & 460 & 67.2 &  0    \\ 
	  & 31.0  &  98.5 & 1.502 &0.546& 26.4 &  0   & 382 & 73.0 &  0    \\ 
	  & 33.0  & 109   & 1.446 &0.592& 28.6 &  0   & 386 & 79.2 &  0    \\ 
	  & 35.06 & 109   & 1.492 &0.536& 30.9 &  0   & 414 & 85.5 &  0    \\ 
	  & 37.12 & 131.5 & 1.412 &0.596& 33.2 &  0   & 436 & 91.8 &  0    \\ 
	  & 39.0  & 128.5 & 1.436 &0.570& 35.3 &  0   & 443 & 97.6 &  0    \\ 
	  & 41.05 &  91.5 & 1.496 &0.566& 37.5 &  0   & 353 & 104 &  0    \\ 
	  & 43.0  &  95.5 & 1.488 &0.562& 39.7 &  0   & 362 & 110 &  0    \\ 
	  & 44.98 &  96   & 1.476 &0.566& 41.9 &  0   & 357 & 116 &  0    \\ 
	  & 46.8  & 109   & 1.468 &0.558& 43.9 & 0  & 398 & 122 & 0   \\ 
$^{53}$Cr & 25.0  & 111.5 & 1.492 &0.552& 19.7 & 3.2 & 425 & 54.3 & 11.4 \\ 
$^{56}$Fe & 25.0  & 102   & 1.512 &0.554& 19.5 & 3.5 & 402 & 53.6 & 12.2 \\ 
          & 26.45 & 108.5 & 1.478 &0.554& 21.1 & 1.4 & 401 & 58.0 & 4.94 \\ 
$^{58}$Fe & 25.0  & 134.5 & 1.390 &0.616& 19.3 & 3.7 & 427 & 53.1 & 12.6 \\ 
$^{59}$Co & 25.0  & 121   & 1.438 &0.584& 19.3 & 3.8 & 417 & 52.9 & 12.8 \\ 
$^{58}$Ni & 8.1   &  70   & 1.858 &0.354&  0.6 &25.9 & 481 & 1.57 & 87.5 \\ 
	  & 9.1   &  79.5 & 1.718 &0.436&  1.7 &25.9 & 502 & 4.63 & 87.5 \\ 
	  & 9.6   &  78.5 & 1.664 &0.466&  2.2 &22.5 & 398 & 6.15 & 76.0 \\ 
	  & 18.0  &  59.5 & 1.862 &0.490& 11.6 &13.9 & 420 & 31.8 & 46.8 \\ 
	  & 21.0  &  77   & 1.670 &0.570& 14.9 & 9.5 & 404 & 40.9 & 32.2 \\ 
	  & 24.1  & 104   & 1.476 &0.574& 18.3 & 5.0 & 385 & 50.4 & 17.0 \\ 
	  & 25.0  & 103.5 & 1.504 &0.558& 19.3 & 3.7 & 402 & 53.1 & 12.6 \\ 
	  & 26.5  & 112.5 & 1.466 &0.558& 21.0 & 1.6 & 406 & 57.7 & 5.29 \\ 
	  & 27.0  & 169.5 & 1.236 &0.662& 21.6 & 0.8 & 398 & 59.2 & 2.85 \\ 
	  & 29.0  & 120.5 & 1.458 &0.556& 23.8 & 0  & 428 & 65.4 & 0   \\ 
	  & 32.3  & 134.5 & 1.412 &0.570& 27.4 & 0  & 439 & 75.4 & 0   \\ 
	  & 34.0  & 209   & 1.426 &0.516& 29.3 & 0  & 688 & 80.6 & 0   \\ 
	  & 37.0  & 140   & 1.384 &0.597& 32.7 & 0  & 436 & 89.8 & 0   \\ 
	  & 38.0  & 139   & 1.400 &0.572& 33.8 & 0  & 443 & 92.8 & 0   \\ 
	  & 43.0  &  99.5 & 1.502 &0.524& 39.3 & 0  & 381 & 108 & 0   \\ 
	  & 49.0  &  73.5 & 1.600 &0.482& 46.0 & 0  & 334 & 126 & 0   \\ 
$^{60}$Ni & 18.0  & 143.5 & 1.522 &0.526& 11.4 &14.1 & 570 & 31.4 & 46.6 \\ 
	  & 21.0  & 136.5 & 1.416 &0.604& 14.8 & 9.7 & 453 & 40.5 & 32.2 \\ 
	  & 24.1  & 129.5 & 1.388 &0.636& 18.2 & 5.2 & 411 & 50.0 & 17.4 \\ 
	  & 25.0  &  99.5 & 1.520 &0.566& 19.2 & 3.9 & 398 & 52.7 & 13.1 \\ 
	  & 27.0  & 139   & 1.364 &0.592& 21.4 & 1.0 & 414 & 58.8 & 3.46 \\ 
	  & 29.0  & 106   & 1.520 &0.506& 23.6 & 0  & 417 & 64.9 & 0   \\ 
	  & 32.3  & 127   & 1.434 &0.560& 27.3 & 0  & 430 & 74.9 & 0   \\ 
	  & 34.0  & 149   & 1.458 &0.540& 29.2 & 0  & 526 & 80.1 & 0   \\ 
$^{62}$Ni & 12.8  & 162   & 1.286 &0.684&  5.5 &21.8 & 425 & 15.2 & 70.8 \\ 
	  & 14.56 & 162.5 & 1.428 &0.574&  7.5 &19.2 & 546 & 20.5 & 62.5 \\ 
	  & 16.34 & 132   & 1.440 &0.586&  9.5 &16.6 & 456 & 25.9 & 54.2 \\ 
	  & 18.0  & 151   & 1.310 &0.652& 11.3 &14.3 & 411 & 31.0 & 46.4 \\ 
	  & 18.13 & 139.5 & 1.378 &0.622& 11.5 &14.1 & 431 & 31.4 & 45.8 \\ 
	  & 21.0  & 127   & 1.454 &0.574& 14.6 & 9.9 & 449 & 40.1 & 32.3 \\ 
	  & 24.1  & 100.5 & 1.508 &0.582& 18.1 & 5.4 & 395 & 49.5 & 17.7 \\ 
	  & 25.0  & 125.5 & 1.412 &0.622& 19.1 & 4.1 & 415 & 52.3 & 13.4 \\ 
	  & 27.0  & 106.5 & 1.500 &0.534& 21.3 & 1.2 & 406 & 58.3 & 4.04 \\ 
	  & 32.3  & 119.5 & 1.444 &0.560& 27.2 & 0  & 412 & 74.5 & 0   \\ 
$^{64}$Ni & 18.0  & 167.5 & 1.224 &0.744& 11.2 &14.4 & 395 & 30.6 & 46.1 \\ 
	  & 21.0  & 112   & 1.524 &0.566& 14.5 &10.1 & 450 & 39.7 & 32.3 \\ 
	  & 24.1  & 115   & 1.462 &0.604& 18.0 & 5.6 & 416 & 49.1 & 18.0 \\ 
	  & 25.0  & 115   & 1.454 &0.596& 19.0 & 4.3 & 408 & 51.8 & 13.8 \\ 
	  & 27.0  &  96.5 & 1.550 &0.534& 21.2 & 1.4 & 404 & 57.9 & 4.58 \\ 
	  & 32.3  & 115.5 & 1.466 &0.548& 27.1 & 0  & 414 & 74.0 & 0   \\ 
$^{63}$Cu & 25.0  & 114.5 & 1.464 &0.580& 19.0 & 4.2 & 413 & 52.1 & 13.6 \\ 
$^{70}$Ge & 25.0  & 103.5 & 1.500 &0.566& 18.6 & 4.9 & 396 & 50.7 & 14.8 \\ 
$^{72}$Ge & 25.0  &  90.5 & 1.556 &0.548& 18.5 & 5.1 & 382 & 50.3 & 15.1 \\ 
$^{74}$Ge & 25.0  & 115.5 & 1.420 &0.614& 18.4 & 5.2 & 382 & 49.9 & 15.3 \\ 
$^{76}$Ge & 25.0  & 115   & 1.428 &0.614& 18.3 & 5.4 & 386 & 49.6 & 15.6 \\ 
$^{76}$Se & 25.0  & 113   & 1.520 &0.592& 18.3 & 5.4 & 450 & 49.6 & 15.6 \\ 
$^{78}$Se & 25.0  & 109.5 & 1.440 &0.602& 18.2 & 5.6 & 375 & 49.2 & 15.8 \\ 
$^{80}$Se & 25.0  & 111.5 & 1.520 &0.578& 18.1 & 5.7 & 441 & 48.9 & 16.1 \\ 
$^{89}$Y  & 16.2  & 133   & 1.404 &0.606&  7.9 &19.2 & 421 & 21.1 & 50.2 \\ 
	  & 19.44 & 136   & 1.356 &0.634& 11.4 &14.5 & 393 & 30.8 & 38.0 \\ 
	  & 21.0  & 122   & 1.438 &0.552& 13.2 &12.2 & 407 & 35.5 & 32.1 \\ 
          & 23.4  & 110.5 & 1.468 &0.574& 15.8 & 8.8 & 394 & 42.7 & 23.0 \\ 
          & 25.0  & 115   & 1.472 &0.540& 17.6 & 6.4 & 409 & 47.4 & 16.9 \\ 
$^{90}$Zr & 15.0  & 167.5 & 1.310 &0.656&  6.5 &21.0 & 443 & 17.4 & 54.6 \\ 
          & 21.0  & 125   & 1.426 &0.566& 13.1 &12.3 & 409 & 35.3 & 32.0 \\ 
          & 23.4  & 107.5 & 1.486 &0.550& 15.8 & 8.8 & 394 & 42.5 & 23.0 \\ 
          & 25.0  & 115   & 1.464 &0.550& 17.6 & 6.5 & 404 & 47.3 & 17.0 \\ 
$^{91}$Zr & 21.0  & 122   & 1.440 &0.558& 13.1 &12.4 & 409 & 35.2 & 32.0 \\ 
          & 23.0  & 105.5 & 1.500 &0.542& 15.3 & 9.5 & 396 & 41.2 & 24.5 \\ 
          & 25.0  & 114.5 & 1.474 &0.536& 17.5 & 6.6 & 409 & 47.1 & 17.1 \\ 
$^{92}$Mo & 13.83 & 188.5 & 1.228 &0.704&  5.1 &22.8 & 423 & 13.7 & 58.6 \\ 
          & 16.42 & 149.5 & 1.372 &0.602& 8.0  &19.1 & 442 & 21.4 & 49.0 \\ 
          & 19.5  & 164.5 & 1.346 &0.610& 11.4 &14.6 & 462 & 30.6 & 37.5 \\ 
$^{94}$Mo & 25.2  & 136.5 & 1.438 &0.534& 17.6 & 6.5 & 453 & 47.3 & 16.5 \\ 
$^{107}$Ag& 15.0  & 117.5 & 1.484 &0.534&  5.7 &22.2 & 425 & 15.3 & 51.7 \\ 
          & 25.2  & 167.5 & 1.252 &0.700& 17.0 & 7.5 & 391 & 45.5 & 17.4 \\ 
$^{112}$Sn& 14.4  & 101.5 & 1.566 &0.476&  4.8 &23.4 & 424 & 12.9 & 52.9 \\ 
	  & 19.5  & 138.5 & 1.420 &0.574& 10.5 &16.0 & 444 & 28.0 & 36.2 \\ 
$^{116}$Sn& 25.2  & 130.5 & 1.442 &0.550& 16.7 & 8.0 & 434 & 44.4 & 17.8 \\ 
$^{122}$Sn& 25.2  & 113.5 & 1.492 &0.524& 16.4 & 8.4 & 414 & 43.7 & 18.0 \\ 
$^{124}$Sn& 19.5  & 130   & 1.424 &0.578& 10.0 &16.8 & 419 & 26.6 & 35.4 \\ 
	  & 25.2  & 101.5 & 1.580 &0.508& 16.3 & 8.5 & 436 & 43.4 & 18.0 \\ 
\hline \hline
\end{longtable}

\newpage

\begin{table*} 
\caption{\label{tab:nomp}Optical potential parameters of the regional potential 
(ROP). The energies are in MeV and geometry parameters in fm.}
\begin{tabular}{llll}\\
\hline \hline
   \hspace*{0.9in} Potential depth  & &\hspace*{0.8in}Geometry parameters\\ 
   \hspace*{1.2in}     (MeV)        & &\hspace*{1.4in}(fm)               \\ 
\hline
V$_R$=168+0.733Z/A$^{1/3}$-2.64E, &E$<$E$_3$ &r$_R$=1.18+0.012E, &E$<$25\\
\hspace*{0.2in}=116.5+0.337Z/A$^{1/3}$-0.453E,&E$\ge$E$_3$&
  \hspace*{0.15in}=1.48,&E$\ge$25\\ 
                                  &          &
  a$_R$=0.671+0.0012A+(0.0094-0.0042A$^{1/3}$)E$_2$,&E$<$E$_2$\\ 
                                  &          &
  \hspace*{0.15in}=max&E$\ge$E$_2$\\ 
  &          &  \hspace*{0.25in}(0.671+0.0012A+(0.0094-0.0042A$^{1/3}$)E,\\    
  &          &  \hspace*{0.30in}0.55),\\
W$_V$=2.73-2.88A$^{1/3}$+1.11E    &          &r$_V$=1.34\\ 
                                  &          &a$_V$=0.50\\ 
W$_D$=4 ,                         &E$<$E$_1$ &r$_D$=1.52\\
\hspace*{0.3in}=22.2+4.57A$^{1/3}$-7.446E$_2$+6E,&E$_1$$<$E$<$E$_2$&
																									a$_D$=0.729-0.074A$^{1/3}$\\        
\hspace*{0.3in}=22.2+4.57A$^{1/3}$-1.446E ,      &E$_2<$E  & \\                                                    
\hline \hline
\end{tabular} 
where:\\
\hspace*{0.0in}E$_1$=-3.03-0.762A$^{1/3}$+1.24E$_2$ 
\hspace*{0.2in}E$_2$=(2.59+10.4/A)Z/R$_B$
\hspace*{0.2in}E$_3$=23.6+0.181Z/A$^{1/3}$\\        
R$_B$=2.66+1.36A$^{1/3}$ \cite{wn80}
\end{table*}

\clearpage
\section*{Figure captions}

\noindent
Fig. 1. (Color online) Comparison of experimental (see Table 1) 
angular distributions of the elastic scattering of $\alpha$-particles 
at 25 MeV on $^{51}$V, $^{53}$Cr, $^{56,58}$Fe, $^{59}$Co, $^{63}$Cu, 
$^{70,74,76}$Ge, $^{76,78,80}$Se, $^{94}$Mo, $^{107}$Ag and 
$^{112,124}$Sn, divided by the Rutherford cross section, with OMP
calculations by using either the present local parameter sets of 
Table 2 (solid curves), the regional parameter set in Table 3 
(dash--dotted curves), or the global OMP parameter set of Kumar et al. 
\cite{ak06} (dotted curves). 

\noindent
Fig. 2. (Color online) The same as in Fig. 1 but for $\alpha$-particles 
scattered on $^{50}$Ti between 25 and 47 MeV.

\noindent
Fig. 3. (Color online) The same as in Fig. 1 but for $\alpha$-particles 
scattered on $^{52}$Cr between 23 and 47 MeV.

\noindent
Fig. 4. (Color online) The same as in Fig. 1 but for $\alpha$-particles 
scattered on $^{50}$Cr between 13 and 25 MeV, and $^{62}$Ni between 13 
and 32 MeV.

\noindent
Fig. 5. (Color online) The same as in Fig. 1 but for $\alpha$-particles
scattered on $^{60}$Ni between 18 and 34 MeV, and $^{64}$Ni between 18 
and 32 MeV.

\noindent
Fig. 6. (Color online) The same as in Fig. 1 but for $\alpha$-particles 
scattered on $^{58}$Ni between 8 and 49 MeV.

\noindent
Fig. 7. (Color online) The same as in Fig. 1 but for $\alpha$-particles 
scattered on $^{89}$Y, $^{90,91}$Zr, $^{94}$Mo, $^{107}$Ag and 
$^{112,124}$Sn, between 14 and 25 MeV.

\noindent
Fig. 8. The $\chi^2$ deviation per degree of freedom between the 
experimental and calculated angular distributions of the elastic 
scattering of $\alpha$-particles on $^{112}$Sn at 14.4 MeV, versus
the real potential depth $V_R$ (solid curve), as well as the values
of the real--potential radius $r_R$, diffuseness $a_R$ and volume 
integral per interacting nucleon pair $J_R$, corresponding to the 
$\chi^2$ minima. The dotted curves are drawn to guide the eye.

\noindent
Fig. 9. The real--potential diffuseness $a_R$ obtained by analysis
of the experimental angular distributions of $\alpha$-particle 
elastic scattering (Table 1) using the average values for the rest 
of all OMP parameters (Table 3), versus the $\alpha$-particle 
energy (top) and its ratio in the center--of--mass system to the 
Coulomb barrier $B_C$ (bottom). The full circles correspond to the 
target nuclei with $A>$89.

\noindent
Fig. 10. (Color online) Comparison of measured \cite{lrg03} angular 
distributions of the elastic scattering of $\alpha$-particles between 
8.1 and 9.6 MeV on $^{58}$Ni, with OMP calculations using (a) the 
local imaginary--potential parameters (Table 2) and semi--microscopic 
DF approach together with the surface--imaginary potential dispersive 
corrections formerly used within the elastic--scattering analysis 
(dotted curves) and corresponding to the final form of the ROP surface imaginary--potential in Table 3 (solid curves), and (b) the
phenomenological parameters obtained formerly within the 
elastic--scattering analysis, i.e. for energies above the limit $E_2$
in Table 3 (dashed curves), and the final ROP (solid curves). In the 
inset of (a) are shown the same surface--imaginary potential dispersive 
corrections used for calculation of the angular distributions shown by
the similar curves.

\noindent
Fig. 11. (Color online) Comparison of measured total $\alpha$-reaction 
cross sections for $^{48}$Ti and $^{51}$V \cite{hv83}, and the related 
$(\alpha,n)$ reaction cross sections for $^{48}$Ti \cite{hv83,ajm92,cmb04}
and $^{51}$V \cite{hv83,aev74,vyh93,xp99}, as well as for 
$^{45}$Sc \cite{aev74,vyh89}, $^{46}$Ti \cite{aev74,ajh74}, 
$^{50}$Cr \cite{aev74,ajm94} and $^{54}$Fe \cite{aev74,sgt91}, and 
$^{44}$Ti$(\alpha,p)^{47}$V reaction cross sections \cite{aas00}, with
calculated values using the predictions of the present optical potential 
established by the elastic--scattering data analysis alone at energies 
above the Coulomb barrier $B_C$ (dash--dotted curves) as well as its 
final form (Table 3) proved necessary for the $(\alpha,x)$ reaction data 
account (solid curves).

\noindent
Fig. 12. (Color online) Comparison of measured $(\alpha,\gamma)$,  
$(\alpha,n)$ and $(\alpha,p)$ reaction cross sections for $^{55}$Mn 
\cite{sgt93},  $^{58}$Ni \cite{aev74,fkmcg64,mr74}, $^{62}$Ni 
\cite{jlz79,mes86}, and $^{64}$Ni \cite{jlz79} target nuclei, 
and calculated values using the predictions of the present optical 
potential established by the elastic--scattering data analysis alone 
at energies above the Coulomb barrier $B_C$ (dash--dotted curves), as 
well as a constant real--potential diffuseness $a_R$ at lowest 
energies below the energy $E_2$ given in Table 3 (dashed curves), and 
the final ROP proved necessary for the $(\alpha,x)$ reaction data 
account (solid curves). The total $\alpha$-reaction cross sections 
provided at all energies by the ROP parameters established by the 
elastic--scattering analysis alone, are also shown (dotted curves) 
for a direct view of the weight of reactions being analyzed, by 
comparison with the corresponding dash--dotted curves.

\noindent
Fig. 13. (Color online) The same as in Fig. 12 but for the target nuclei 
$^{56}$Fe \cite{mra79}, $^{59}$Co \cite{phs64,jmda68,oaz73,fmm75,sgt88},
$^{63}$Cu \cite{msb05,phs64,fmm75,ntp59,jz87}, 
$^{65}$Cu \cite{phs64,fmm75,ntp59,nls94}, and $^{70}$Ge \cite{zf96}.

\noindent
Fig. 14. (Color online) The same as in Fig. 12 but for the target 
nuclei $^{96,98}$Ru \cite{wr02}.

\noindent
Fig. 15. (Color online) The same as in Fig. 12 but for the target 
nuclei $^{106}$Cd \cite{gg06} and $^{112,118}$Sn \cite{no07,sh05},
except the bottom--right corner where the final ROP proved necessary 
for the $(\alpha,x)$ reaction data account is used for all calculated
cross sections of $(\alpha,n)$ (thick solid curve), $(\alpha,\gamma)$
(thin solid curve), $(\alpha,\alpha')$ (short--dotted curve), and
$(\alpha,p)$ (dash--dot--dotted curve) reactions.

\noindent
Fig. 16. (Color online) The energies $E_1$ (dash--dotted curve) below
which the imaginary--potential depth $W_D$=4 MeV, $E_2$ (solid curve) 
corresponding to 0.9$B_C$, and $E_3$ (dotted curve), at which the 
present ROP parameters change their energy dependences, versus 
the target nuclei atomic--mass number, and the energy ranges of the 
$(\alpha,x)$ reaction data formerly analyzed in the present work 
(thin vertical bars) as well as involved within the additional check
of ROP (thick bars).
The mass--dependences corresponding to nuclei with a nuclear asymmetry 
(N-Z)/A value of 0.1 are shown, while the complete formulas of the 
energies $E_1$, $E_2$, $E_3$ are given in Table 3.

\newpage

\noindent
\begin{figure*}
\resizebox{17.5cm}{!}{
  \includegraphics{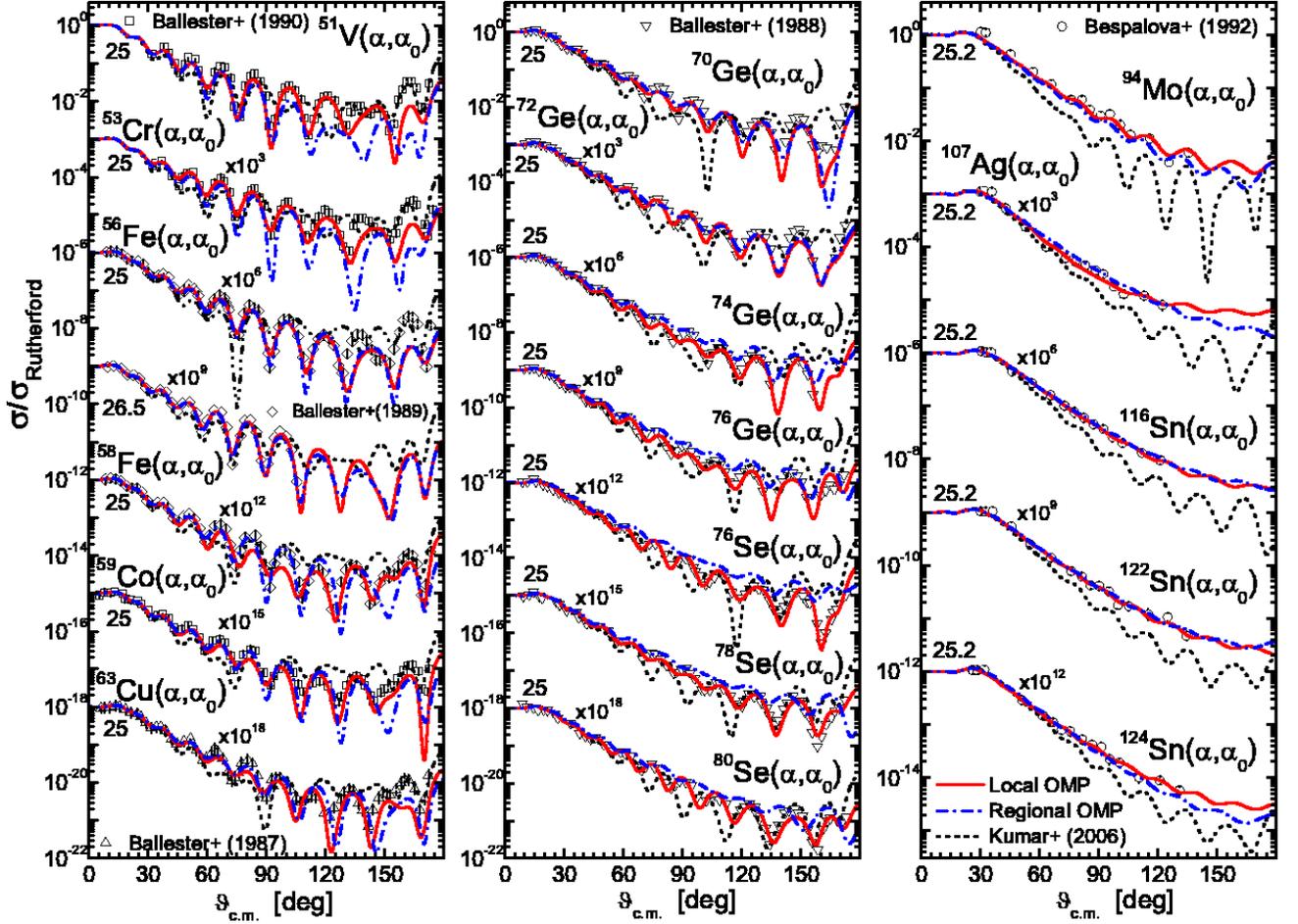}}%
\caption{\label{Fig1}(Color online) Comparison of experimental (see 
Table 1) angular distributions of the elastic scattering of 
$\alpha$-particles at 25 MeV on $^{51}$V, $^{53}$Cr, $^{56,58}$Fe, 
$^{59}$Co, $^{63}$Cu, $^{70,74,76}$Ge, $^{76,78,80}$Se, $^{94}$Mo, 
$^{107}$Ag and $^{112,124}$Sn, divided by the Rutherford cross section, 
with OMP calculations by using either the present local parameter 
sets of Table 2 (solid curves), the regional parameter set in Table 3 
(dash--dotted curves), or the global OMP parameter set of Kumar et al. 
\cite{ak06} (dotted curves).}
\end{figure*}

\begin{figure*}
\resizebox{17.5cm}{!}{
  \includegraphics{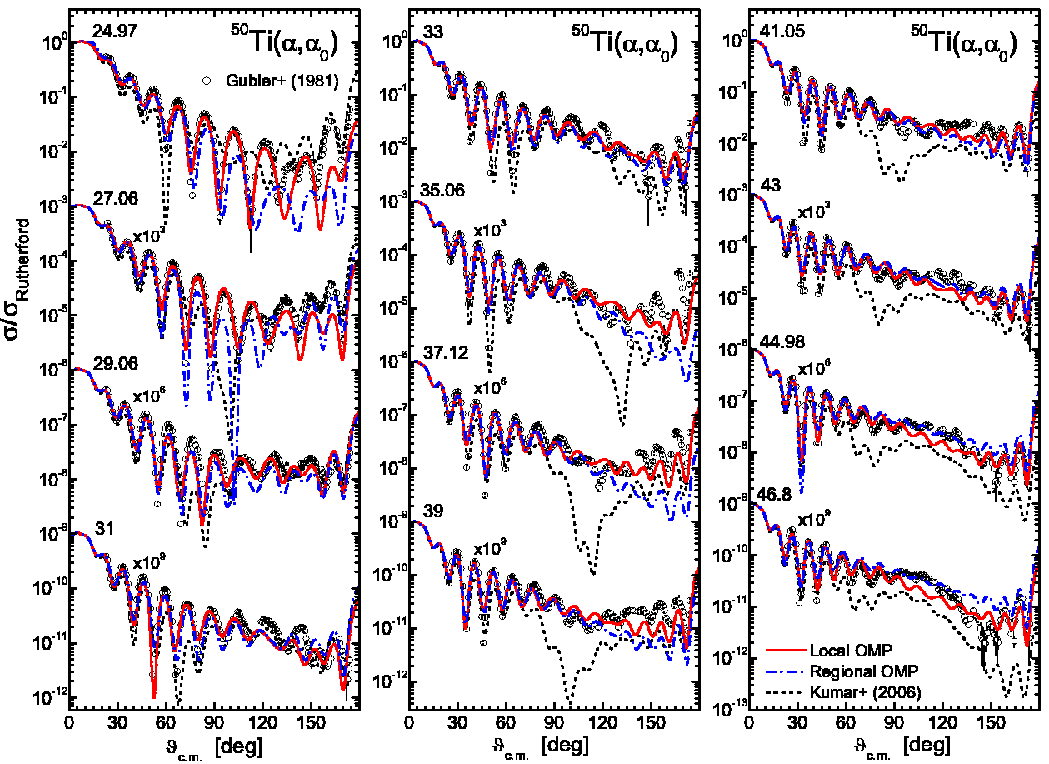}}%
\caption{\label{Fig2}(Color online) The same as in Fig. 1 but for 
$\alpha$-particles scattered on $^{50}$Ti between 25 and 47 MeV.}
\end{figure*}

\begin{figure*}
\resizebox{17.5cm}{!}{
  \includegraphics{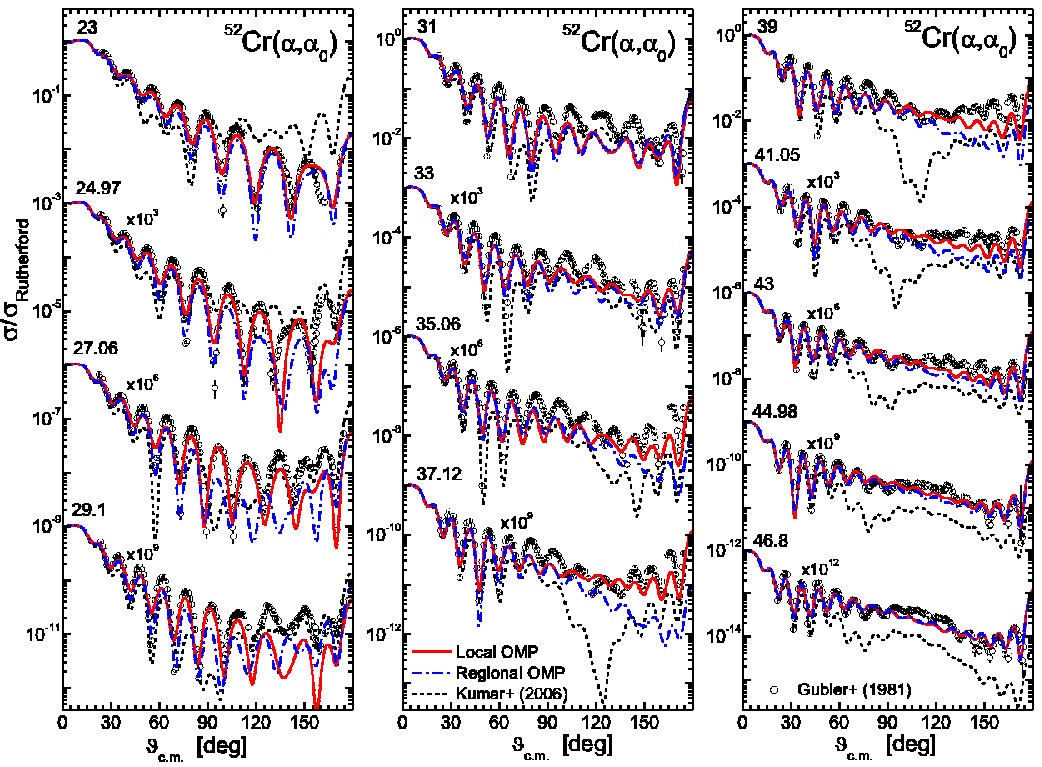}}%
\caption{\label{Fig3}(Color online) The same as in Fig. 1 but for 
$\alpha$-particles scattered on $^{52}$Cr between 23 and 47 MeV}
\end{figure*}

\begin{figure*}
\resizebox{17.5cm}{!}{
  \includegraphics{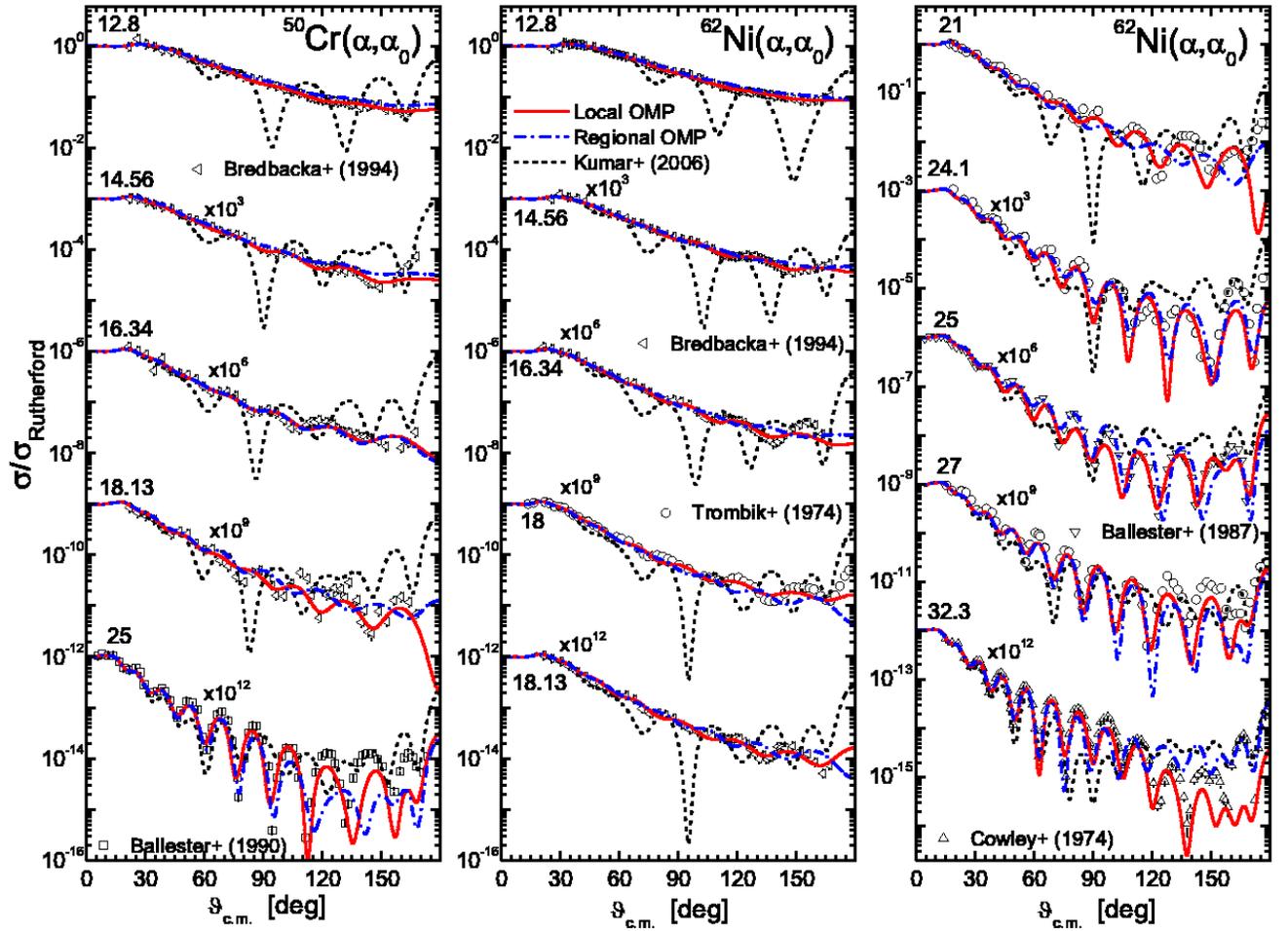}}%
\caption{\label{Fig4}(Color online) The same as in Fig. 1 but for 
$\alpha$-particles scattered on $^{50}$Cr between 13 and 25 MeV, 
and $^{62}$Ni between 13 and 32 MeV.}
\end{figure*}

\begin{figure*}
\resizebox{17.5cm}{!}{
  \includegraphics{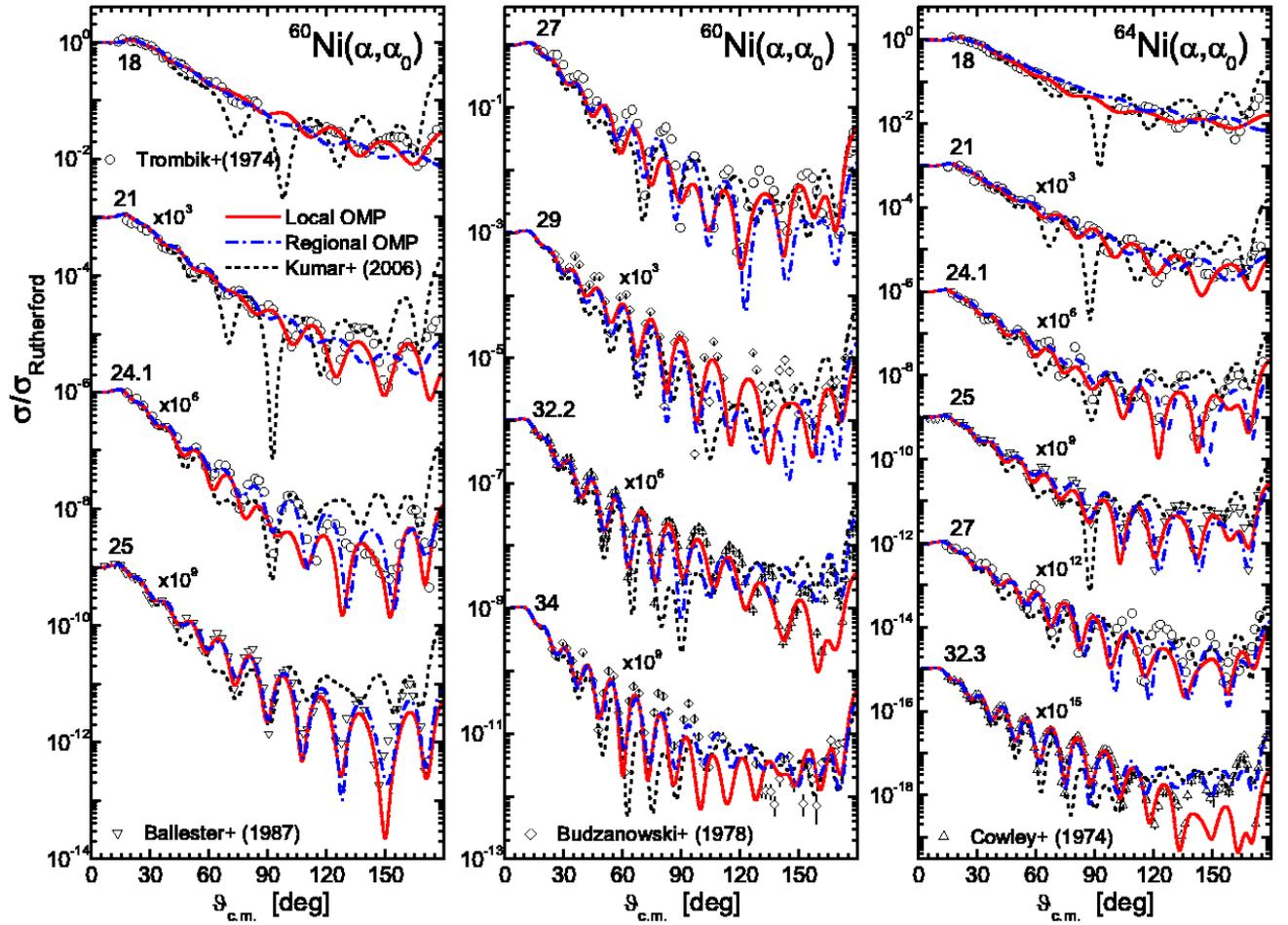}}%
\caption{\label{Fig5}(Color online) The same as in Fig. 1 but for 
$\alpha$-particles scattered on $^{60}$Ni between 18 and 34 MeV, 
and $^{64}$Ni between 18 and 32 MeV.}
\end{figure*}

\begin{figure*}
\resizebox{17.5cm}{!}{
  \includegraphics{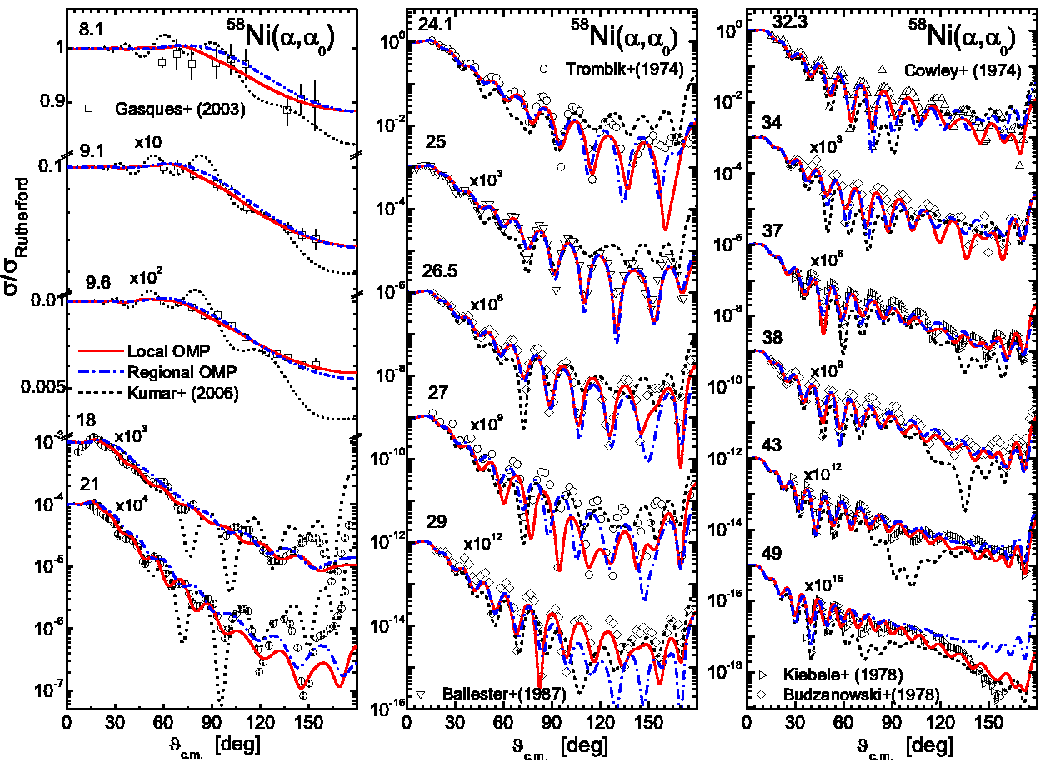}}%
\caption{\label{Fig6}(Color online) The same as in Fig. 1 but for 
$\alpha$-particles scattered on $^{58}$Ni between 8 and 49 MeV.}
\end{figure*}

\begin{figure*}
\resizebox{17.5cm}{!}{
  \includegraphics{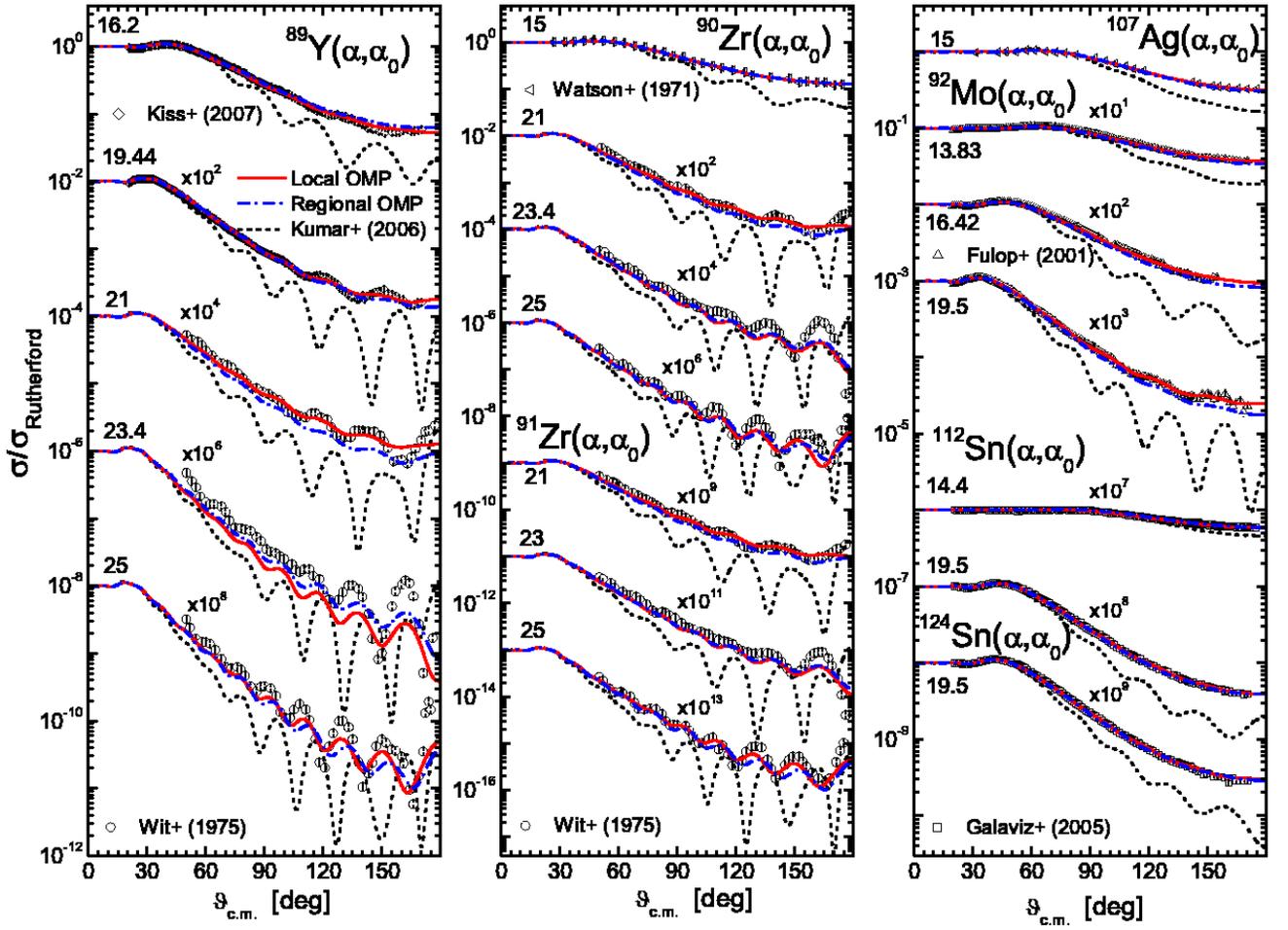}}%
\caption{\label{Fig7}(Color online) The same as in Fig. 1 but for 
$\alpha$-particles scattered at 21, 23.4 and 25 MeV on $^{89}$Y, 
$^{90,91}$Zr, $^{94}$Mo, $^{107}$Ag and $^{112,124}$Sn, between 
14 and 25 MeV.}
\end{figure*}

\begin{figure*}
\resizebox{5.6cm}{!}{
  \includegraphics{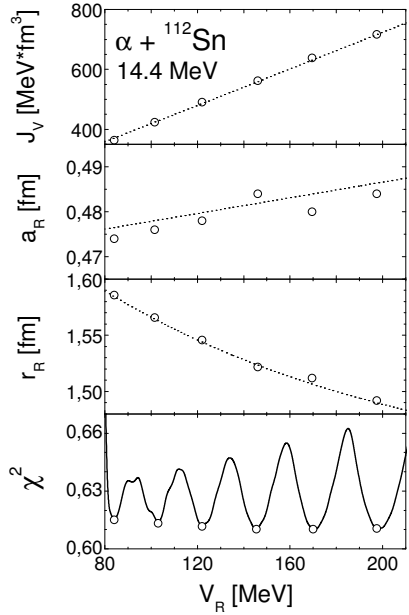}}%
\caption{\label{Fig8} The $\chi^2$ deviation per degree of freedom 
between the experimental and calculated angular distributions of the 
elastic scattering of $\alpha$-particles on $^{112}$Sn at 14.4 MeV, 
versus the real potential depth $V_R$ (solid curve), as well as the 
values of the real--potential radius $r_R$, diffuseness $a_R$ and 
volume integral per interacting nucleon pair $J_R$, corresponding 
to the $\chi^2$ minima. The dotted curves are drawn to guide the eye.}
\end{figure*}

\begin{figure*}
\resizebox{5.6cm}{!}{
  \includegraphics{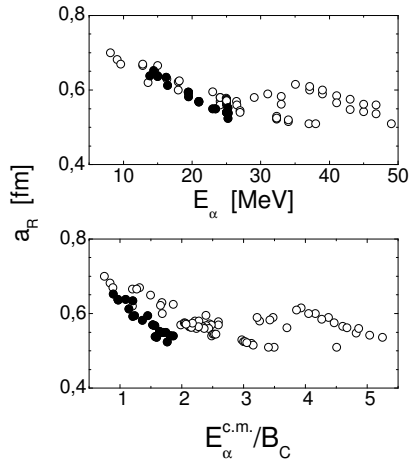}}%
\caption{\label{Fig9} The real--potential diffuseness $a_R$ obtained 
by analysis of the experimental angular distributions of 
$\alpha$-particle elastic scattering (Table 1) using the average 
values for the rest of all OMP parameters (Table 3), versus the 
$\alpha$-particle energy (top) and its ratio in the center--of--mass 
system to the Coulomb barrier $B_C$ (bottom). The full circles 
correspond to the target nuclei with $A>$89.}
\end{figure*}

\begin{figure*}
\resizebox{11.5cm}{!}{
  \includegraphics{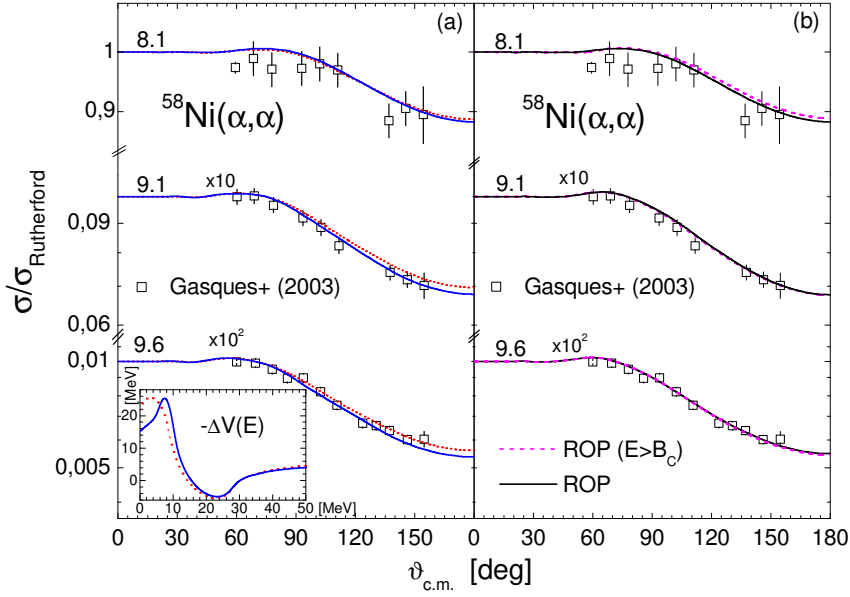}}%
\caption{\label{Fig10} (Color online) Comparison of measured \cite{lrg03} 
angular distributions of the elastic scattering of $\alpha$-particles 
between 8.1 and 9.6 MeV on $^{58}$Ni, with OMP calculations using (a) the 
local imaginary--potential parameters (Table 2) and semi--microscopic 
DF approach together with the surface--imaginary potential dispersive 
corrections formerly used within the elastic--scattering analysis 
(dotted curves) and corresponding to the final form of the ROP surface imaginary--potential in Table 3 (solid curves), and (b) the
phenomenological parameters obtained formerly within the 
elastic--scattering analysis, i.e. for energies above the limit $E_2$
in Table 3 (dashed curves), and the final ROP (solid curves). In the 
inset of (a) are shown the same surface--imaginary potential dispersive 
corrections used for calculation of the angular distributions shown by
the similar curves.}
\end{figure*}

\begin{figure*}
\resizebox{14.cm}{!}{
  \includegraphics{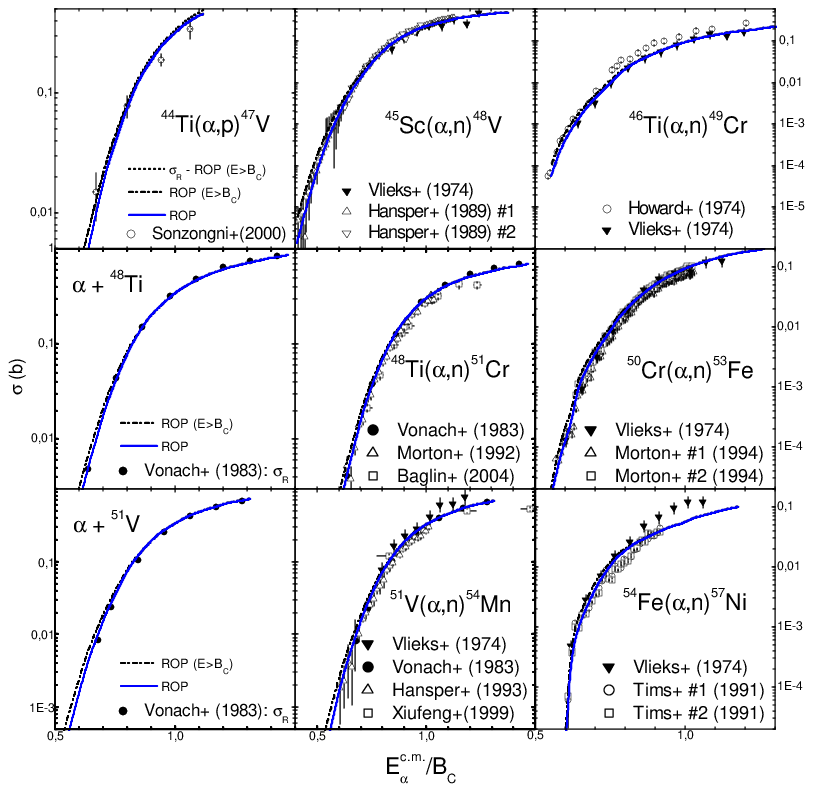}}%
\caption{\label{Fig11}Color online) Comparison of measured total $\alpha$-reaction 
cross sections for $^{48}$Ti and $^{51}$V \cite{hv83}, and the related 
$(\alpha,n)$ reaction cross sections for $^{48}$Ti \cite{hv83,ajm92,cmb04}
and $^{51}$V \cite{hv83,aev74,vyh93,xp99}, as well as for 
$^{45}$Sc \cite{aev74,vyh89}, $^{46}$Ti \cite{aev74,ajh74}, 
$^{50}$Cr \cite{aev74,ajm94} and $^{54}$Fe \cite{aev74,sgt91}, and 
$^{44}$Ti$(\alpha,p)^{47}$V reaction cross sections \cite{aas00}, with
calculated values using the predictions of the present optical potential 
established by the elastic--scattering data analysis alone at energies 
above the Coulomb barrier $B_C$ (dash--dotted curves) as well as its 
final form (Table 3) proved necessary for the $(\alpha,x)$ reaction data 
account (solid curves).}
\end{figure*}

\begin{figure*}
\resizebox{13.8cm}{!}{
  \includegraphics{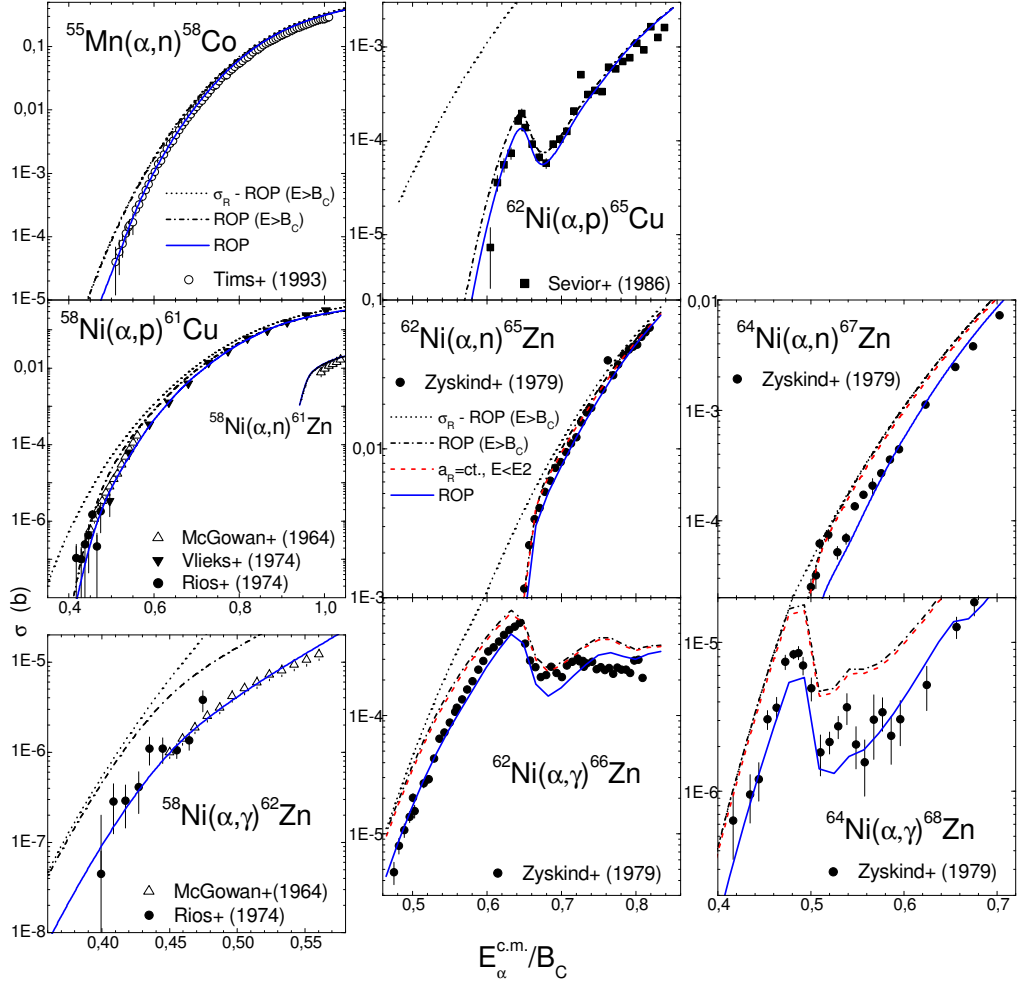}}%
\caption{\label{Fig12}(Color online) Comparison of measured 
$(\alpha,\gamma)$, $(\alpha,n)$ and $(\alpha,p)$ reaction cross 
sections for $^{55}$Mn \cite{sgt93}, $^{58}$Ni \cite{aev74,fkmcg64,mr74}, 
$^{62}$Ni \cite{jlz79,mes86}, and $^{64}$Ni \cite{jlz79} target nuclei, 
and calculated values using the predictions of the present optical 
potential established by the elastic--scattering data analysis alone 
at energies above the Coulomb barrier $B_C$ (dash--dotted curves), as 
well as a constant real--potential diffuseness $a_R$ at lowest 
energies below the energy $E_2$ given in Table 3 (dashed curves), and 
the final ROP proved necessary for the $(\alpha,x)$ reaction data 
account (solid curves). The total $\alpha$-reaction cross sections 
provided at all energies by the ROP parameters established by the 
elastic--scattering analysis alone, are also shown (dotted curves) 
for a direct view of the weight of reactions being analyzed, by 
comparison with the corresponding dash--dotted curves.}
\end{figure*}

\begin{figure*}
\resizebox{14.cm}{!}{
  \includegraphics{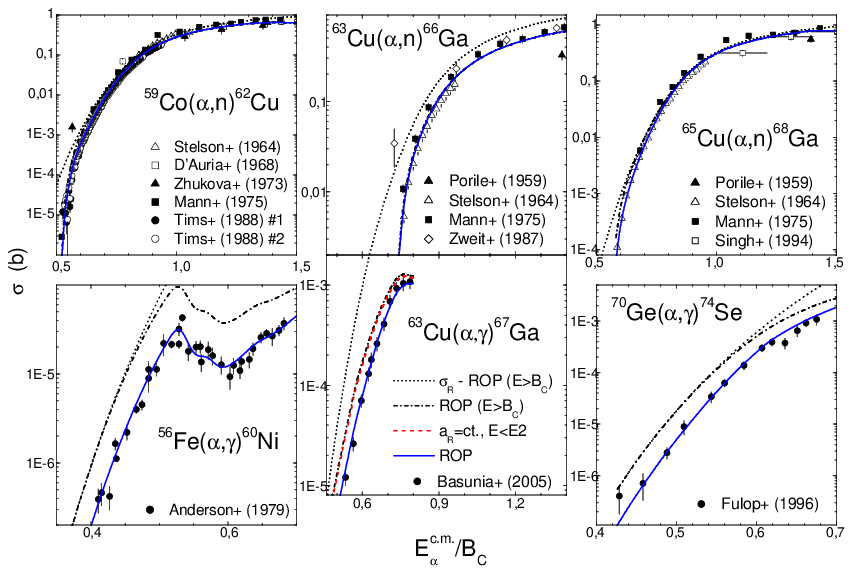}}%
\caption{\label{Fig13}(Color online) The same as in Fig. 12 but for 
the target nuclei $^{56}$Fe \cite{mra79}, 
$^{59}$Co \cite{phs64,jmda68,oaz73,fmm75,sgt88},
$^{63}$Cu \cite{msb05,phs64,fmm75,ntp59,jz87}, 
$^{65}$Cu \cite{phs64,fmm75,ntp59,nls94}, and $^{70}$Ge \cite{zf96}.}
\end{figure*}

\begin{figure*}
\resizebox{10.cm}{!}{
  \includegraphics{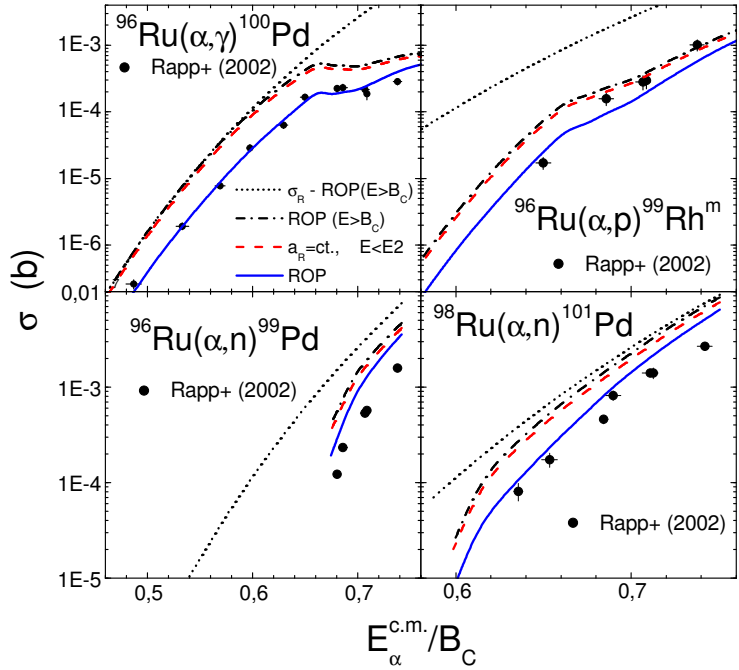}}%
\caption{\label{Fig14}(Color online) The same as in Fig. 12 but 
for the target nuclei $^{96,98}$Ru \cite{wr02}.}
\end{figure*}

\begin{figure*}
\resizebox{10.cm}{!}{
  \includegraphics{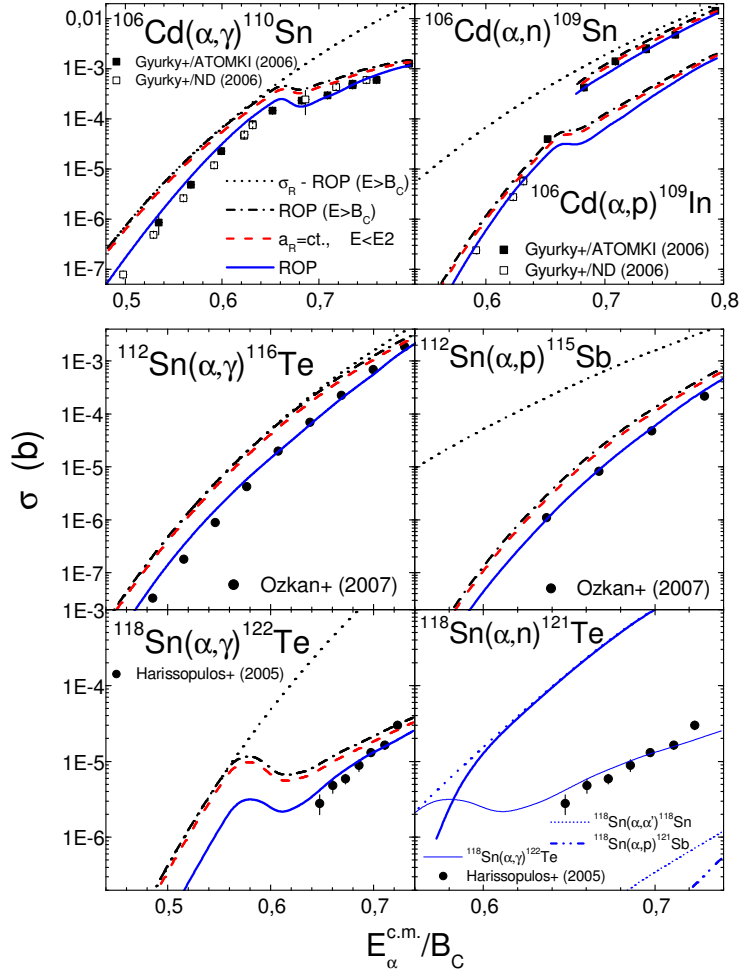}}%
\caption{\label{Fig15}(Color online) The same as in Fig. 12 but 
for the target nuclei $^{106}$Cd \cite{gg06} and $^{112,118}$Sn 
\cite{no07,sh05}, except the bottom--right corner where the final 
ROP proved necessary for the $(\alpha,x)$ reaction data account is 
used for all calculated cross sections of $(\alpha,n)$ (thick solid 
curve), $(\alpha,\gamma)$ (thin solid curve), $(\alpha,\alpha')$ 
(short--dotted curve), and $(\alpha,p)$ (dash--dot--dotted curve) 
reactions.}
\end{figure*}

\begin{figure*}
\resizebox{8.0cm}{!}{
  \includegraphics{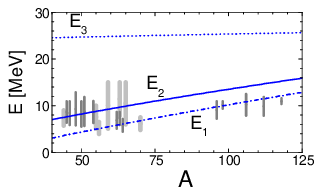}}%
\caption{\label{Fig16}(Color online) The energies $E_1$ (dash--dotted curve) 
below which the imaginary--potential depth $W_D$=4 MeV, $E_2$ (solid curve) 
corresponding to 0.9$B_C$, and $E_3$ (dotted curve), at which the 
present ROP parameters change their energy dependences, versus 
the target nuclei atomic--mass number, and the energy ranges of the 
$(\alpha,x)$ reaction data formerly analyzed in the present work 
(thin vertical bars) as well as involved within the additional check
of ROP (thick bars).
The mass--dependences corresponding to nuclei with a nuclear asymmetry 
(N-Z)/A value of 0.1 are shown, while the complete formulas of the 
energies $E_1$, $E_2$, $E_3$ are given in Table 3.}
\end{figure*}


\begin{thebibliography}{99}
\bibitem{peh90} P.E. Hodgson, Contemp. Phys. 31 (1990) 99; 
	31 (1990) 295.
\bibitem{pm06} P.~Mohr, Phys. Rev. C 73 (2006) 031301; 
	V.~Yu.~Denisov, H.~Ikezoe, Phys. Rev. C 72 (2005) 064613.
\bibitem{mar07} M. Arnould, S. Goriely, K. Takahashi, 
		Phys. Rep. 450 (2007) 97.
\bibitem{mn87} M. Nolte, H. Machner, J. Bojowals, 
	Phys. Rev. C 36 (1987) 1312.
\bibitem{lwp74} L.W. Put,  A.M.J. Paans, Phys. Lett. B 49 (1974) 266;
        Nucl. Phys. A291 (1977) 93.
\bibitem{pps76} P.P. Singh,  P. Schwandt, Nukleonika 21 (1976) 451.
\bibitem{jma90} J.M. Alexander, M.T. Magda, S. Landowne,
        Phys. Rev. C 42 (1990) 1092.
\bibitem{va94} V. Avrigeanu, P.E. Hodgson, M. Avrigeanu, 
	Phys. Rev. C 49 (1994) 2136.
\bibitem{glr87} G. La Rana et al., 
Phys. Rev. C 35 (1987) 373; 
G. D.J. Moses et al., 
Phys. Rev. C 36 (1987) 422;
R. Lacey et al., Phys. Lett. B 191 (1987) 253;
G. Bozzolo, O. Civitarese, J.P. Vary, Phys. Lett. B 219 (1989) 161.
\bibitem{ma06a} M. Avrigeanu, W. von Oertzen, V. Avrigeanu, 
	Nucl. Phys. A 764 (2006) 246.
\bibitem{pd02} P. Demetriou, C. Grama, S. Goriely, 
	Nucl. Phys. A 707 (2002) 253.	
\bibitem{ma03} M. Avrigeanu, W. von Oertzen, A.J.M. Plompen,
	 V. Avrigeanu, Nucl. Phys. A 723 (2003) 104.
\bibitem{ma07a} M. Avrigeanu, W. von Oertzen, A. Obreja, F.L. Roman, 
	V. Avrigeanu, in Proc. of Int. Conf. on Nuclear Data for Science
	and Technology, Nice, France (2007) (in press).
\bibitem{gg06} Gy. Gyürky, G.G. Kiss, Z. Elekes, Zs. Fülöp, 
	E. Samorjai, A. Palumbo, J. G\"orres, H.Y. Lee, W. Rapp, M. Wiescher,
	N. \"Ozkan, R.T. G\"uray, G. Efe,  T. Rauscher,
	Phys. Rev. C 74 (2006) 025805.
\bibitem{no07} N. \"Ozkan, G. Efe, R.T. G\"uray, A. Palumbo, 
	J. G\"orres, H.Y. Lee, L.O. Lamm, W. Rapp, E. Stech, M. Wiescher, 
	Gy. Gyürky, Zs. Fülöp, 	E. Samorjai, 
	Phys. Rev. C 75 (2007) 025801.
\bibitem{lmf66} L. McFadden, G.R. Satchler, Nucl. Phys. A84 (1966) 177.
\bibitem{tr03} T. Rauscher, Nucl. Phys. A 719 (2002) 73c; A725	(2003) 295(E).
\bibitem{pek04}	P.E. Koehler, Yu.M. Gledenov, T. Rauscher, 
	C. Fr\"ohlich, Phys. Rev. C 69 (2004) 015803.
\bibitem{dg05} D. Galaviz, Zs. F\"ul\"op, Gy. Gy\"urky, Z. M\'at\'e, 
	P. Mohr, T. Rauscher, E. Somorjai, A. Zilges,
	Phys. Rev. C 71 (2005) 065802.
\bibitem{msb05} M.S. Basunia, E.B. Norman, H.A. Shugart, A.R. Smith, 
	 M.J. Dolinski, B.J. Quiter, Phys. Rev. C 71 (2005) 035801.
\bibitem{wr02} W. Rapp, M. Heil, D. Hentschel, R. Reifarth, H.J. Brede,
	H. Klein, T. Rauscher, Phys. Rev. C 66 (2002) 015803.
\bibitem{sh05} S. Harissopulos, A. Lagoyannis, A. Spyrou, Ch., Zarkadas,
	S. Galanopoulos, G. Perdikakis, H.-W. Becker, C. Rolfs, F. Strieder,
	R. Kunz, M. Fey, J.W. Hammer, A. Dewald, K.-O. Zell, P. von Brentano,
	R. Julin, P. Demetriou, 
	J. Phys. G: Nucl. Part. Phys. 31 (2005) S1417.
\bibitem{ak06} A. Kumar, S. Kailas, S. Rathi, K. Mahata, 
	Nucl. Phys. A 776 (2006) 105.	
\bibitem{ua96} U. Atzrott, P. Mohr, H. Abele, C. Hillenmayer, 
	G. Staudt, Phys. Rev. C 53 (1996) 1336.
\bibitem{zf01} Zs. F\"ul\"op, Gy. Gy\"urky, Z. M\'at\'e, E. Somorjai, 
	L. Zolnai, D. Galaviz, M. Babilon, P. Mohr, A. Zilges, T. Rauscher,
	H. Oberhummer, G. Staudt, Phys. Rev. C 64 (2001) 065805.
\bibitem{ggk07} G.G. Kiss, Zs. Fülöp, Gy. Gyürky, Z. Màtè, E. Samorjai, 
	D. Galaviz, S. M\"uller, A. Zilges, P. Mohr, M. Avrigeanu, 
	J. Phys. G: Nucl. Part. Phys. 35 (2008), 
	doi:10.1088/0954-3899/35/1/014037.
\bibitem{ma06b} M. Avrigeanu, V. Avrigeanu, 
	Phys. Rev. C 73 (2006) 038801.
\bibitem{hv83} H. Vonach, R.C. Haight, G. Winkler, 
	Phys. Rev. C 28 (1983) 2278.
\bibitem{pr01} P. Reimer, V. Avrigeanu, A.J.M. Plompen, S.M. Qaim,
        Phys. Rev. C 65 (2001) 014604.
\bibitem{vs04} V. Semkova, V. Avrigeanu, T. Glodariu, A.J. Koning, 
	A.J.M. Plompen, D.L. Smith, S. Sudar, Nucl. Phys. A 730 (2004) 255.
\bibitem{ma07b} M. Avrigeanu, S.V. Chuvaev, A.A. Filatenkov, 
	R.A. Forrest, M. Herman, A.J. Koning, A.J.M. Plompen, F.L. Roman,
	V. Avrigeanu, arXiv:0712.0699 [nucl--ex]; Nucl. Phys. A 608 (2008) 15.
\bibitem{pr05} P. Reimer, V. Avrigeanu, S. Chuvaev, A.A. Filatenkov,
	T. Glodariu, A.J. Koning, A.J.M. Plompen, S.M. Qaim, D.L. Smith,
	H. Weigmann, Phys. Rev. C 71 (2005) 044617.
\bibitem{va07} V. Avrigeanu, F.L. Roman, M. Avrigeanu, in: A. Plompen 
	(Ed.), 	Proc. 4th NEMEA-4 Workshop on Neutron Measurements, 
	Evaluations and Applications, Prague, Czechia (2007),	European 
	Commission Report EUR 23235 EN, Belgium, 2008, p. 143. Available 
	from: www.irmm.jrc.be/html/publications/.
\bibitem{ma07c} M. Avrigeanu, V. Avrigeanu, A. Obreja, F.L. Roman,
	OECD/NEA Data Bank Report EFFDOC--1022, Nov. 2007.
\bibitem{grs79} G.R. Satchler, W.G. Love, Phys. Rep. 55 (1979) 183.
\bibitem{meb97} M.E. Brandan, G.R. Satchler, Phys. Rep. 285 (1997) 143.
\bibitem{dtk01} Dao T. Khoa, Phys. Rev. C 63 (2001) 034007.
\bibitem{gb77} G. Bertsch, J. Borysowicz, H. McManus, W.G. Love, 
	Nucl. Phys. A 284 (1977) 399.
\bibitem{na83} N. Anantaraman, H. Toki, G. Bertsch, 
	Nucl. Phys. A 398 (1983) 279.
\bibitem{dtk00} Dao T. Khoa, G.R. Satchler, 
	Nucl. Phys. A668 (2000) 3.	
\bibitem{dtk93} Dao T. Khoa, W. von Oertzen, 
	Phys. Lett. B 304 (1993) 8; 342 (1995) 6; 
	Dao T. Khoa, W. von Oertzen, H.G. Bohlen,
	Phys. Rev. C 49 (1994) 1652; Dao T. Khoa, W. von Oertzen,
	A.A. Ogloblin, Nucl. Phys. A 602 (1996) 98.
\bibitem{dtk97} D.T. Khoa, G.R. Satchler, W. von Oertzen, 
        Phys. Rev. C 56 (1997) 954.
\bibitem{db96} D. Baye, L. Desorgher, D. Guillain, D. Herschkowitz,
	Phys. Rev. C 54 (1996) 2563.
\bibitem{it92} I. Tanihata, D. Hirata, T. Kobayashi, S. Shimoura, 
	K. Sugimoto, H. Toki, Preprint RIKEN--AF--NP--123, 1992; 
	Phys. Lett. B 289 (1992) 261.
\bibitem{jwn70} J.W. Negele, Phys. Rev. C 1 (1970) 1260.
\bibitem{jwl76} J.W. Lightbody, Jr., S. Penner, S.P. Fivozinsky, 
	P.L. Hallowell, H. Crannell, Phys. Rev. C 14 (1976) 952.
\bibitem{mef85} M. El--Azab Farid, G.R. Satchler, 
	Nucl. Phys. A 438 (1985) 525.
\bibitem{amk82} A.M. Kobos, B.A. Brown, P.E. Hodgson, G.R. Satchler, 
	 A. Budzanowski, Nucl. Phys. A 384 (1982) 65.
\bibitem{ha93} H. Abele, G. Staudt, Phys. Rev. C 47 (1993) 742.
\bibitem{ai00} A. Ingemarsson, J. Nyberg, P.U. Renberg, O. Sundberg,
	R.F. Calrson, A.J. Cox, A. Auce, R. Johansson, G. Tibell, 
	Dao T. Khoa, R.E. Warner, Nucl. Phys. A 676 (2000) 3. 
\bibitem{cm86a} C. Mahaux, H. Ngo, G.R. Satchler, 
	Nucl. Phys. A 449 (1986) 354.
\bibitem{cm86b} C. Mahaux, H. Ngo, G.R. Satchler, 
	Nucl. Phys. A 456 (1986) 134.
\bibitem{grs91} G.R. Satchler, Phys. Rep. 199 (1991) 147.
\bibitem{hpg81} H.P. Gubler, U. Kiebele, H.O. Meyer, G.R. Plattner, 
	I. Sick, Nucl. Phys. A 351 (1981) 29; EXFOR-D0257 data file 
	entry, EXFOR Nuclear Reaction Data, http://www-nds.iaea.or.at/exfor .
\bibitem{fb90} F. Ballester, E. Casal, J.B.A. England, 
	Nucl. Phys. A 513 (1990) 61; EXFOR-01087 data file entry.
\bibitem{ab94} A. Bredbacka, M. Brenner, K.M. Kallman, P. Manngard, 
	Z. Mate, S. Szilagyi, L. Zolnai, 
	Nucl. Phys. A 574 (1994) 397; EXFOR-F0461 data file entry.
\bibitem{fb89} F. Ballester, E. Casal, J.B.A. England, 
	Nucl. Phys. A 501 (1989) 301; EXFOR-F0528 data file entry.
\bibitem{bw62} A.G. Blair, H.E. Wegner, Phys. Rev. 127 (1962) 1233;
        EXFOR-C1022 data file entry.
\bibitem{fb87} F. Ballester, E. Casal, J. Diaz, J.B.A. England,
	F. Moriano, J. Phys. G 13 (1987) 1541; EXFOR-01089 data file entry.
\bibitem{lrg03} L.R. Gasques et al., 
	Phys. Rev. C 67 (2003) 024602; EXFOR-C1165 data file entry.
\bibitem{wt74} W. Trombik, K.A. Eberhard, O. Hinderer, H.H. Rossner, 
	A. Weidinger, J.S. Eck, Phys. Rev. C 9 (1974) 1813; 
	EXFOR-D0372 data file entry. 
\bibitem{ab78} A.Budzanowski et al.,
	Phys. Rev. C 17 (1978) 951; EXFOR-D0296 data file entry.
\bibitem{uk78} U. Kiebele, E. Baumgartner, H.P. Gubler, H.O. Meyer, 
	G.R. Plattner, I. Sick, Helvetica Physica Acta 51 (1978) 726; 
        EXFOR-D0373 data file entry.
\bibitem{aac74} A.A. Cowley, P.M. Cronje, G. Heymann, S.J. Mills, 
	J.C.van Staden, Nucl. Phys. A 229 (1974) 256;
	EXFOR-D0374 data file entry.
\bibitem{fb88g} F. Ballester, E. Casal, J.B.A. England, F. Moriano,
	 Nucl. Phys. A 490 (1988) 227; EXFOR-F0077 data file entry.
\bibitem{fb88s} F. Ballester, E. Casal, J.B.A. England, F. Moriano,
	J. Phys. G 14 (1988) 1103; EXFOR-O1088 data file entry.
\bibitem{wit75} M. Wit, J. Schiele, K.A. Eberhard, J.P. Schiffer,
        Phys. Rev. C 12 (1975) 1447; EXFOR-D0345 data file entry.
\bibitem{watson71} B.D. Watson, D. Robson, D.D. Tolbert, R.H. Davis, 
	Phys. Rev. C 4 (1971) 2240.
\bibitem{bespalova92} O.V. Bespalova, E.A. Romanovskij, N.G. Gorjaga,
	Nguen Mak Kha, B.S. Galakhmatova, Luaj Morzena Rafu, S.I. Fedoseev, 
	Dang Lam, Anis Belal, Yad. Fiz. 56 (1992) 113; 
        EXFOR-A0545 data file entry.
\bibitem{scat2} O. Bersillon, Centre d'Etudes de Bruyeres--le--Chatel,
        Note CEA--N--2227, 1992.		
\bibitem{pm97} P. Mohr, T. Rauscher, H. Oberhummer, Z. M\'at\'e,
	Zs. F\"ul\"op, E. Somorjai, M. Jaeger, G. Staudt, 
	Phys. Rev. C 55 (1997) 1523.
\bibitem{rmd63} R.M. Drisko, G.R. Satchler, R.H. Bassel, 
	Phys. Lett. 5 (1963) 347.
\bibitem{wn80}W. N\"orenberg, in: Heavy Ion Collisions, R. Bock (Ed.),
	North--Holland, Amsterdam, 1980.
\bibitem{pm00} P.~Mohr, Phys. Rev. C 61 (2000) 045802.
\bibitem{ma95} M. Avrigeanu, V. Avrigeanu, Report NP--86--1995,
	Bucharest, IPNE, 1995; News NEA Data Bank 17 (1995) 22.
\bibitem{geb81}G.E. Brown, M. Rho, Nucl. Phys. A 372 (1981) 397.
\bibitem{es98} E. Somorjai, Zs. F\"ul\"op, A.Z. Kiss, C.E. Rolfs, 
	H.-P. Trautvetter, U. Greife, M. Junker, S. Goriely, M. Arnould,
	M. Rayet, T. Rauscher, H. Oberhummer, Astron. Astrophys. 333 (1998) 1112.
\bibitem{ajm92} A.J. Morton, S.G. Tims, A.F. Scott, V.Y. Hansper, 
	C.I.W. Tingwell, D.G. Sargood, Nucl. Phys. A 537 (1992) 167.
\bibitem{cmb04} C.M. Baglin, E.R. Norman, R.M. Larimer, G.A. Rech, 
	AIP Conf. Proc. 769 (2005) 1370.
\bibitem{aev74} A.E. Vlieks, J.F. Morgan, S.L. Blatt, 
	Nucl. Phys. A 224 (1974) 492.
\bibitem{vyh93} V.Y. Hansper, A.J. Morton, S.G. Tims, C.I.W. Tingwell, 
	A.F. Scott, D.G. Sargood, Nucl. Phys. A 551 (1993) 158.
\bibitem{xp99} Xiufeng Peng, Fuqing He, Xianguan Long, 
	Nucl. Inst. Meth. B 152 (1999) 432.
\bibitem{vyh89} V.Y. Hansper, C.I.W. Tingwell, S.G. Tims, A.F. Scott, 
	D.G. Sargood, Nucl. Phys. A 504 (1989) 605.
\bibitem{ajh74} A.J. Howard, H.B. Jensen, M. Rios, W.A. Fowler, 
	B.A. Zimmerman, Astrophys. J. 188 (1974) 131; EXFOR-C0180 data file entry.	
\bibitem{ajm94} A.J. Morton, A.F. Scott, S.G. Tims, V.Y. Hansper, 
	D.G. Sargood, Nucl. Phys. A 573 (1994) 276.
\bibitem{sgt91} S.G. Tims, A.J. Morton, C.I.W. Tingwell, A.F. Scott, 
	V.Y. Hansper, D.G. Sargood, Nucl. Phys. A 524 (1991) 479.
\bibitem{aas00} A.A. Sonzongni et al., Phys. Rev. Lett. 84 (2000) 1651.
\bibitem{sgt93} S.G. Tims, A.F. Scott, A.J. Morton, V.Y. Hansper, 
	D.G. Sargood, Nucl. Phys. A 563 (1993) 473.
\bibitem{fkmcg64} F.K. McGowan, P.H. Stelson, W.G. Smith,
	Phys. Rev. 133 (1964) B907.
\bibitem{mr74} M. Rios, B.D. Anderson, J.S. Schweitzer, 
	Nucl. Phys. A 236 (1974) 523.
\bibitem{jlz79} J.L. Zyskind, J.M. Davidson, M.T. Esat, M.H. Shapiro,
	R.H. Spear, Nucl. Phys. A 331 (1979) 180.
\bibitem{mes86} M.E. Sevior, L.M. Mitchell, C.I.W. Tingwell, D.G. Sargood, 
	Nucl. Phys. A 454 (1986) 128.
\bibitem{mra79}	M.R. Anderson, S.R. Kennett, Z.E. Switkowski, D.G. Sargood,
	Nucl. Phys. A 318 (1979) 471.
\bibitem{phs64} P.H. Stelson, F.K. McGowan, Phys. Rev. 133 (1964) B911.
\bibitem{jmda68} J.M. D'Auria, M.J. Fluss, L. Kowalski, J.M. Miller,
	Phys. Rev. 168 (1968) 1224.
\bibitem{oaz73} O.A. Zhukova, V.I. Kanashevich, S.V. Laptev, G.P. Chursin,
	Sov. J. Nucl. Phys. 16 (1973)	134; EXFOR-A0644 data file entry.
\bibitem{fmm75} F.M. Mann, R.W. Kavanach, Nucl. Phys. A 255 (1975) 287.
\bibitem{sgt88} S.G.Tims, C.I.W. Tingwell, V.Y. Hansper, A.F. Scott, 
	D.G.Sargood, Nucl. Phys. A 483 (1988) 354.
\bibitem{ntp59} N.T. Porile, D.L. Morrison, Phys. Rev. 116 (1959) 1193.
\bibitem{jz87} J. Zweit, H. Sharma, S. Downey, 	
	Appl. Radiat. Isot. 38 (1987) 499; EXFOR-D0119 data file entry.
\bibitem{nls94} N.L. Singh, B.J. Patel, D.R.S. Somayajulu, S.N. Chintalapudi, 
	Pramana 42 (1994) 349; EXFOR-D0099 data file entry.
\bibitem{zf96} Zs. Fulop, A.Z. Kiss, E. Somorjai, C.E. Rolfs, H.P. Trautvetter, 
	T. Rauscher, H. Oberhummer, Z. Phys. A 355 (1996) 203; EXFOR-O0897 data file 
	entry.
\end{thebibliography}
\end{document}